\documentclass[11pt]{article}
\usepackage[utf8]{inputenc}
\usepackage[T1]{fontenc}
\usepackage{graphicx}
\usepackage{grffile}
\usepackage{longtable}
\usepackage{wrapfig}
\usepackage{rotating}
\usepackage[normalem]{ulem}
\usepackage{amsmath}
\usepackage{textcomp}
\usepackage{amssymb}
\usepackage{capt-of}
\usepackage{hyperref}

\newcommand{\close}{c^{\textrm{close}} }
\newcommand{\bet}{c^{\textrm{betw}} }

\newcommand{\ev}{c^{\textrm{ev}} }

\newcommand{\nSamples}{\num{1000}}

\newcommand{\step}{\num{10000}}

\newcommand{\vo}{\vec{o}\@ifnextchar{^}{\,}{}}

\newcommand{\codebase}{\href{https://cvanelteren.github.io}{cvanelteren.github.io} }

\usepackage{siunitx} %
\usepackage{mathtools}
\usepackage{amsmath} %
\usepackage{cleveref}%
\usepackage{tabularx}
\usepackage{MnSymbol}%
\usepackage{wasysym}%

\usepackage{float} %
\usepackage{tabularx} %

\usepackage{subcaption}
\usepackage{multirow} %
\usepackage{booktabs} %
\usepackage{wasysym}%

\usepackage{graphicx}
\usepackage{multirow, multicol} %
\usepackage[final]{pdfpages}

\usepackage{xspace}
\usepackage{listings}
\usepackage{arxiv}
\usepackage{csvsimple}
\usepackage[utf8]{inputenc} %
\usepackage{url}            %
\usepackage{booktabs}       %
\usepackage{amsfonts}       %
\usepackage{nicefrac}       %
\usepackage{microtype}      %

\usepackage{lipsum}
\usepackage[title]{appendix}
\usepackage[font={scriptsize,it}]{caption}
\usepackage{algorithm2e}
\usepackage{algorithmic}
\usepackage{amssymb,amsmath,amsthm}

\SetKwFor{For}{for (}{) $\lbrace$}{$\rbrace$}

\newtheorem{Definition}{Definition}
\newtheorem{Corollary}{Corollary}
\newtheorem{theorem}{Theorem}

\usepackage{grffile} %

\usepackage[backend = biber, sorting = nty, style = nature, url = false, isbn = false]{biblatex} %

\bibliography{library.bib}
\DeclareSourcemap{
	\maps[datatype = biblatex]{
		\map[overwrite = true]{
			\step[fieldset = abstract, null]
		}
	}
}

\makeatletter
\renewcommand{\@algocf@capt@plain}{above}%
\makeatother

\author{
 Casper van Elteren \\
 \\ %
 \\ %
 caspervanelteren@gmail.com
 \AND Rick Quax \\
 r.quax@uva.nl\\
 \AND Peter Sloot \\
 p.m.a.sloot@uva.nl\\
}

\usepackage[frozencache=true]{minted}
\date{\today}
\title{Dynamic importance of network nodes is poorly predicted by static structural features}
\hypersetup{
 pdfauthor={Casper van Elteren},
 pdftitle={Dynamic importance of network nodes is poorly predicted by static structural features},
 pdfkeywords={},
 pdfsubject={},
 pdfcreator={Emacs 27.2.50 (Org mode 9.5)}, 
 pdflang={English}}
\begin{document}

\maketitle
\begin{abstract}
One of the  most central questions in  network science is:
which  nodes are  most important?  Often this  question is
answered   using  structural   properties  such   as  high
connectedness  or  centrality  in  the  network.  However,
static  structural  connectedness   does  not  necessarily
translate to dynamical importance. To demonstrate this, we
simulate  the  kinetic  Ising   spin  model  on  generated
networks and one real-world  weighted network. The dynamic
impact  of nodes  is assessed  by causally  intervening on
node state  probabilities and measuring the  effect on the
systemic  dynamics.  The   results  show  that  structural
features such  as network centrality or  connectedness are
actually poor predictors of the dynamical impact of a node
on the rest  of the network. A solution is  offered in the
form   of  an   information   theoretical  measure   named
integrated  mutual  information.  The metric  is  able  to
accurately  predict the  dynamically  most important  node
(``driver'' node) in networks based on observational data of
non-intervened  dynamics.  We  conclude  that  the  driver
node(s)  in   networks  are   not  necessarily   the  most
well-connected  or  central   nodes.  Indeed,  the  common
assumption   of   network    structural   features   being
proportional   to    dynamical   importance    is   false.
Consequently,  great care  should be  taken when  deriving
dynamical  importance  from   network  data  alone.  These
results  highlight the  need for  novel inference  methods
that take both structure and dynamics into account.
\end{abstract}
\section{Introduction}
\label{sec:orgd0d9843}
Understanding complex  systems is a fundamental  problem for
the  21st  century  \cite{Mercury2000}. Despite  the  apparent
differences  and  purposes   of  many  real-world  networked
complex systems,  previous research shows  several universal
characteristics   in  networks   properties   such  as   the
small-world   phenomenon  \cite{Watts1998},   fat-tail  degree
\cite{Albert2002},  and feedback  loops \cite{Thomas2005}.  This
has lead  to the common  but often implicit  assumption that
the connectedness of  a node in the  network is proportional
to its  dynamic importance \cite{Ianishi2020}. For  example in
epidemic  research, high  degree nodes  or ``super-spreaders''
are  associated  to  dominant epidemic  risk  and  therefore
deserve   special  attention   \cite{Sikic2013}.  Yet,   prior
research shows  that the  shared network  characteristics is
not shared in the dynamic  or functional properties that are
exerted  on these  networks \cite{Barzel2013,Harush2017a}.  In
particular, the  dynamic importance  of a  node varies  as a
function of both  the dynamics that exist on  the network in
addition to  its structural  connectedness. This  effect was
rigorously shown by  Harush and colleagues \cite{Harush2017a}.
In  their study,  the  dynamic processes  were varied  while
keeping  the   network  structure  the  same.   Both  random
generated  networks  as  well as  real-world  networks  were
studied.  Nodal   importance  was  computed  based   on  the
``information flow'' through  a node. The flow  through a node
had   a   non-linear   relation   between   its   structural
connectedness  and  the  type  of dynamics  present  in  the
system.  These  results   highlight  the  non-intuitive  and
non-trivial  interplay  of the  structure  of  a system  and
dynamics between the nodes of a system. However, the results
required full knowledge of the system, i.e. both the network
structure and  the dynamics  of the  system are  required to
estimate the node with the highest dynamic importance. For a
real-world system having both the dynamics available and the
underlying network structure may  be difficult. In addition,
in their study the network structure was deemed constant. It
remains unclear  whether the proposed flow  would generalize
to systems with varying  network structure. In addition, the
results  were  obtained  from  steady-state.  For  dynamical
processes  an  out-of-equilibrium approach  highlight  \emph{how}
trajectories generate systemic behavior.  What is needed are
methods that can detect the  highest dynamic importance of a
node from observations directly  without knowing or assuming
the underlying network or causal structure among variables.

A node's  dynamic (or  causal) importance,  is traditionally
inferred  by  means  of  interventions  and  counterfactuals
\cite{Pearl2000,Woodward2015}. Through  external interventions
the  behavior  of  the  system   may  change.  That  is,  an
intervention on a part of  the system causes a divergence of
the system behavior proportional  to its dynamic importance;
high  causal importance  relates to  a large  change in  the
system  behavior.  It is  common  for  studies using  causal
interventions apply  so-called hard  intervention strategies
\cite{Schamberg2020}. In  hard interventions a node  is pinned
to   a  state.   This  effectively   removes  all   incoming
connections to  a node,  removing all influences  the system
has on this node. The use of hard interventions is common in
various  disciplines  such   as  gene-knockout  experiments,
epidemic  spreading,   network  analysis  and   driver  node
identification.

Applying external interventions to  identify nodes with high
dynamical  importance,  referred  to  as  driver  nodes,  is
challenging for three reasons. First, hard interventions may
cause a change  in systemic behavior that does  not occur in
the non-intervened behavior  (e.g. see fig. \ref{fig:methods1}
B  )  or  may  be impossible  in  practice  (e.g.  requiring
infinite resources  \cite{Wang2016}). This  complicates theory
forming  on  what   mechanisms  underlie  observed  systemic
behavior.  Second,  most  approaches  for  causal  inference
assume  that the  causal  interactions  follow a  particular
structure           without          (local)           loops
\cite{Mooij2011,Forre2020,Dablander2019}.    This   assumption
simplifies  theoretical  analysis  and   is  in  many  cases
justified  under the  assumption that  causes precede  their
effects  in time.  However, this  approach assumes  that the
underlying causal  structure is known and  or accessible. If
the underlying causal  structure is not known, it  has to be
inferred  from  data  which  prompts problems  in  terms  of
temporal  resolution  and scale.  For  example  in a  causal
process  where \(A  \to B  \to  C \to  A\), the  cycle may  be
unrolled  over   time.  This  effectively   removes  cycles.
However, this  requires the data to  support time resolution
such  that   it  reflects  unrolled   (non-cyclical)  causal
processes  \cite{Mooij2011}.  Cyclical  causal  structures  do
occur in  real-world systems such as  ecosystems, biological
systems, gene-regulatory systems and so on. Importantly, for
many of these systems determining the correct temporal scale
for  causal interactions  is  non-trivial. Consequently,  it
remains an open question on  how to perform causal inference
applying  these  in  general  for  cyclical  causal  events.
Thirdly, the underlying causal structure may not be known or
difficult to  determine \cite{Epskamp2018}. In  addition, many
dynamical systems are  prohibited from analytical approaches
to decompose each nodal dynamical importance directly due to
the     polyadic,     often     non-linear,     interactions
\cite{Ladyman2013}.

For closed systems, it is possible to avoid these challenges
to  deduce the  driver node  by considering  cross-sectional
time series and computing the  correlation of each node with
the  entire  system  out-of-equilibrium  and  without  lossy
compression  \cite{Stokes2017}. Here,  a  driver  node is  the
element of the  system that is correlated the  most with the
entire  system as  a function  of time.  It has  the maximum
dynamic  (causal) impact  when  intervened  upon across  all
nodes.  By  taking  the  entire  system  into  account,  any
internal  confounding  information is  implicitly  accounted
for; there  cannot exist  any other  node with  more dynamic
importance than the node with the highest correlation as the
system is closed and external confounding is excluded.

Shannon information  theory offers profound  advantages over
previous  approaches for  defining  a  measure of  dynamical
importance of driver node on the behavior of all other nodes
\cite{Cover2005}. Firstly and most prominently, Shannon mutual
information  can  quantify  statistical  associations  among
variables  without bias  to  specific  forms of  association
\cite{Kinney2014}. In  particular, it equals zero  if and only
if  the full  probability distribution  of the  system state
remains  exactly unchanged  regardless of  the state  of the
node. Second, mutual information does not require \emph{a priori}
knowledge of  the representational base of  the system. That
is, it allows for  direct comparison among different systems
that  may  have  different  units  of  measurement  such  as
currency, density  of animals, voltage per  surface area and
so on. Finally,  it is defined for  both discrete-valued and
real-valued state variables.

The concept  of measuring the  dynamic importance of  a node
through   information  flow   is   not  new.   Colloquially,
information flows from process \(X\) to process \(Y\) represents
the existence  of statistical coherence between  the present
information in \(Y\) and the past  of \(X\) not accounted for by
the past of  \(Y\). Various methods have been  proposed in the
past  such as  transfer entropy  \cite{Schreiber2000} and  its
derivatives
\cite{Lizier2013,Sensoy2014,Songhorzadeh2016,Wibral2014}.
Although originally  intended as  a predictive  measure, the
notions of information and  information flow can be extended
to causal influence or dynamic importance. Previous research
developed several  measures and methods for  determining how
much information flow between two processes is truly causal;
examples  include (but  not limited  to) conditional  mutual
information under causal intervention \cite{Ay2008}, causation
entropy   \cite{Sun2014},   permutation   conditional   mutual
information  \cite{Runge2019},   time-delayed  Shannon  mutual
information \cite{Li2018}.

These measures  are commonly  used to infer  the information
transfer between sets of nodes  by possibly correcting for a
third  confounding variable  \cite{Janzing2013,Schamberg2020}.
That  is,  informational flows  are  used  to determine  how
information is  transferred or shared between  pair or sets
of variables.  However, in polyadic settings most measures of
information flow are prone  to underestimate or overestimate
nodal  importance   \cite{James2016}.  Determining   how  much
information flow is causal between source and sink variables
in polyadic settings remains  difficult due to the so-called
synergetic         and         redundant         information
\cite{James2016,Quax2017}.

Instead of focusing on the  open problem of full information
decomposition                 among                variables
\cite{Sun2014,Ay2008,Schamberg2020,Schreiber2000,Janzing2013},
we  focus here  on the  amount  of information  that a  node
shares with the entire system.  A driver node is expected to
have a  corresponding high information  flow from it  to the
system. We introduce a  novel metric named integrated mutual
information  (IMI)  based  on  time-delayed  Shannon  mutual
information with a node and the entire system over time that
captures driver  nodes with  the highest \emph{causal}  impact in
ergodic  systems.  Additionally,  we avoid  the  problem  of
confounding information synergy  with redundancy by avoiding
conditional mutual  information. The consequence of  this is
that we quantify  only \emph{how much} causal  impact \emph{one single
node} has on  the \emph{entire} network. In other  words, we will
not be able to infer which parts of the network are impacted
more  than  others; nor  how  exactly  the causal  influence
percolates  through   the  network.  This   is  nevertheless
sufficient for the purpose of this study.

Using  this approach  it was  previously shown  analytically
that the  number of connections  of a node  does necessarily
not  scale monotonically  with its  dynamical importance  in
infinite-sized,  locally-tree-like  networks (i.e.  networks
without loops) \cite{Quax2013}. For those networks, nodes with
high  dynamic importance  were  not nodes  with high  degree
(so-called  hubs).  Instead,  the  nodes  with  intermediate
connectedness  were identified  with  the highest  dynamical
importance.  This  study  extends  this  prior  research  by
numerically  computing  information  flow  in  finite-sized,
random networks,  i.e. where small feedback  loops cannot be
ignored, and  studies the relation between  connectivity and
the dynamical importance of nodes in such networks.

The aim of this paper is to test the common (often implicit)
hypothesis that the connectedness  of a node is proportional
to its dynamical importance.  A node's dynamic importance is
determined by  simulating the  (out-of-equilibrium) dynamics
of  a  system  under   causal  external  interventions.  The
distance between the  system state probability distributions
with \emph{and} without performing the  intervention is used as a
ground truth  for a node's  causal influence over  time (see
\ref{sec:org98c8d2e}). The resulting
causal  impact  score  over  time is  compared  with  common
network centrality  metrics (\textbf{appendix}  \ref{sec:org9e3ed34})
as well as our proposed information-based metric.

In contrast to  other studies, the proposed  metrics in this
study  (section  \ref{sec:org8af9ae7})  does  not  make
assumptions on network structure  or type of dynamics. Small
networks are used  (up to 12 nodes) for which  each node has
an associated discrete state whose dynamics is governed by a
stochastic  update rule  in discrete  time. Smaller  network
sizes  have  the  of  studying  ``network  motifs''  that  are
embedded   in   larger  network   structure   \cite{Alon2007};
understanding causal flows of smaller structures may provide
insights into larger systems  consisting of a combination of
smaller network  structures. In  addition, the use  of small
network  sizes offers  a  numerical  advantage for  accurate
estimation of  causal and information measures.  In total 16
Erdös-Rényi  networks  are  generated (10  nodes),  and  one
real-world  weighted   network  (12  nodes)   obtained  from
\cite{Fried2015}. Temporal dynamics of the nodes are simulated
using kinetic Ising spin  dynamics with Glauber updating for
the purpose of demonstrating  a case of non-trivial relation
between network connectivity and dynamic importance, but our
approach  easily  generalizes  to   any  other  dynamics  or
networks generation model.

The results show  that nodes with generally  nodes with high
structural   connectedness   as   measured   by   closeness,
betweenness, eigenvector  and degree centrality are  not the
\emph{driver   node(s)}.   Our   novel  metric,   IMI,   achieved
significantly  better performance  in predicting  the driver
node for  non-intervened dynamical systems. In  addition and
most importantly,  hard causal interventions lead  to causal
flows that differ from the non-intervened dynamics. That is,
as a  function of  external intervention,  systemic behavior
may result  in in  ``unnatural'' system  behavior that  do not
occur  in the  unperturbed system.  The proposed  metric IMI
does  not  rely  on  the assumptions  on  dynamics,  nor  on
assumptions  on   structural  properties  of   the  network.
Therefore, the  results of this study  provide scientists of
all fields  a novel,  reliable and  accurate metric  for the
identification of driver nodes.

\section{Theoretical background}
\label{sec:org8af9ae7}
\subsection{Terminology}
\label{sec:org254cd9f}
In this  paper, we  consider a  complex system  as a  set of
discrete  random variables  \(S =  \{s_1, s_2,  \dots, s_n\}\)
with interaction structure \(E  = \{(s_i, s_k) | s_i,
s_k \in S\}\), where each \(s_i  \in S\) has an alphabet \(A\). This
is   also   known   as  a   (discrete)   dynamical   network
\cite{Feng2007a}. Please  note that  we use  the term  node and
variable interchangeably  referring to  an element  \(s_i \in
S\). The system chooses its next state \(S^t\) in discrete time
with probability:

\begin{equation}
\begin{split}
p(S^t | S^{t-1}, \dots, S^{t_0}) = p(S^{t} | S^{t-1}),
\end{split}
\end{equation}

which  is also  known as  a first-order  Markov chain.  More
 specifically, each  discrete time  step, a  single variable
 \(s_i \in S\) is chosen with uniform probability and updated.
 That is,

\begin{equation}
\label{eq:markov}
\begin{split}
p(S^t | S^{t-1}, \dots, S^{t_0}) = p(S^{t} | S^{t-1}) = \prod_j p(s_j^{t} | S^{t-1}) = p(s_i^{t} | S^{t-1}).
\end{split}
\end{equation}

For temporal dynamics, we  adopt here the Metropolis-Hasting
algorithm  \cite{Hastings1970}. By  drawing  a proposal  state
\(S^{t+1} = X'\) from current state  \(S^t = X\) from a proposal
distribution \(g(X'  | X)\) and  accepting the new  state \(X'\)
with probability,

\begin{equation}
\begin{split}
A(S^{t+1} = X', S^t = X) &= \min(1, \frac{p(S^{t+1} = X') g(S^{t+1} = X | S^t = X')}{p(S^t = X) g(S^{t + 1} = X' | S^t = X)}\\
&= \min(1, \frac{p(s_i^{t+1} = x') g(s_i^{t} = x | s_i^{t} = x')}{p(s_i^{t+1} = x) g(s_i^{t} = x' | s_i^{t} = x)}),
\end{split}
\end{equation}
where \(x' \in A\). If the new state is not accepted, then the
next state will  be set to \(S^{t+1} = X\).  As the next state
\(S^{t+1}\)  is  determined  through  considering  updating  a
single variable  \(s_i \in  S\), the next  state \(X'\)  will be
generated through  the proposal  state \(x'\)  drawn uniformly
from the  possible states \(A\) such  that \(g(x' | x)  = g(x |
x') = g(x') = \frac{1}{|A|}\).  This means that \(\frac{g(X' |
X)}{g(X | X')} = \frac{g(x' | x)}{g(x | x')} = 1\).

For our experiments, we use the kinetic Ising model which is
thought to  fall in the  same universality class  as various
other  complex behaviors  \cite{Odor2004}, such  as (directed)
percolation,  diffusion, and  many extensions  of the  model
have  been used  as a  base for  opinion dynamics,  modeling
neural  behavior and  so on.  For demonstrating  our primary
claim that  high network  connectivity does  not necessarily
lead  to  high  dynamical   importance,  a  single  dynamics
suffices as counterexample.

It is nevertheless important  to emphasize that our proposed
driver node  inference method does  not depend on  the exact
type of dynamics. That is,  the choice for the kinetic Ising
spin dynamics is arbitrary in that respect. Our methods only
require  that  for  a  given  dynamics  the  data-processing
inequality is satisfied. More details on the methods and its
assumption will follow in \ref{sec:org1ac74b5}.

The  kinetic  Ising  model   consists  of  binary  variables
dictated  by  a  Gibbs distribution  that  interact  through
nearest neighbor  interactions. A prominent property  of the
Ising model in higher dimensions  (two or more) is the phase
transition from  an ordered phase  to a disordered  phase by
increasing the noise parameter  \(\beta\). For finite systems,
the kinetic Ising model  shows a continuous phase transition
from    ordered   to    unordered   system    regime   (fig.
\ref{fig:methods2}  B).  The Metropolis-Hastings  update  rule
specifically for the kinetic Ising model equals:

\begin{equation}
\label{eq:hastings}
\begin{split}
p(  \text{accept } X' ) = \frac{p(X')}{p(X)} =
\begin{cases}
 1 & \text{if }  \mathcal{H}(X') - \mathcal{H}(X) < 0\\
\exp(-\beta (\mathcal{H}(X') - \mathcal{H}(X)) & \text{otherwise,}
\end{cases}
\end{split}
\end{equation}

where \(\mathcal{H}(S)\) is the system Hamiltonian defined as

\begin{equation}
\label{eq:energy}
\begin{split}
    \mathcal{H}(S) = -\sum_{i,j} J_{ij} s_{i} s_{j} - h_{i} s_{i}.
\end{split}
\end{equation}

Here \(\beta\)  is the  inverse temperature  \(\frac{1}{k_b T}\)
with Boltzmann  constant \(k_b\), \(J_{ij}\) is  the interaction
strength between variables \(s_i\) and \(s_j\); \(h_i\) represents
external influence  on node \(i\). The  matrix \(J\) effectively
represents the network of the  system: an edge between \(s_i,
s_j \in S\) exists if \(|J_{ij}| > 0\). The networks considered
in this study are undirected  networks, which means that \(J\)
is an  \(|S| \times |S|\)  symmetric matrix. For  the randomly
generated networks the edges \(J_{ij}  \in \{0, 1\}\); for the
real  network  dataset the  edge  weights  are positive  and
negative real numbers (section \ref{sec:org4457d3a}).

The \(\beta\)  parameter can  be seen  as the  (inverse) noise
parameter in the  system. Low values of  \(\beta\) will induce
each node in the system to  detach from the influence of its
neighbors, i.e. the probability of finding a node in a state
\(p(s_i = a),  a \in A\) will tend to  uniform distribution as
\(\beta \rightarrow  0\). In contrast, high  values of \(\beta\)
increases the  influence a  neighbor of a  node may  have on
determining the node's next state.

\subsection{Causal interventions and dynamic importance}
\label{sec:org98c8d2e}
We call a  node a \emph{driver node} for a  dynamical system when
it has the largest causal  impact on the system dynamics. In
brief,  we   will  determine  a  node's   causal  impact  by
simulating   a  transient   intervention  and   subsequently
quantifying the difference between the system dynamics under
intervention and  without intervention.  We expect  that the
impact of  the driver nodes  will penetrate deeper  into the
system and remain  present longer than for  nodes with lower
causal impact.  Here, we  define causal  impact by  means of
external  intervention  on  a   node.  The  external  causal
intervention  \(\vec{\epsilon}\) on  node \(s_j  \in S\)  can be
described as

\begin{equation}
\label{eq:intervention}
\begin{split}
p_{s_j}'(s_i^t | S^{t-1}) = p(s_i^t | S^{t-1}) + \vec{\epsilon}\delta_{ij},
\end{split}
\end{equation}

where    \(\delta_{ij}\)   is    the   Kronecker-delta,    and
\(\textrm{dim}(\vec{\epsilon}) = |A|\).  In addition, \(\sum_{i
=    0}^{|A|}   \epsilon_i    =    0\)    and   \(\sum_{i    =
0}^{|A|} |\epsilon_{i}| =  c\) for some \(c \in  (0, 1]\). Note
only those \(\vec{\epsilon}\) are  allowed that generate valid
new probabilities, i.e. \(0 \leq p' \leq 1\).

Relative to some equilibrium distribution \(p(S^{\tau})\), the
effect of intervention \(\vec{\epsilon}\) will result in a new
system   state   equilibrium  distribution   \(p'(S^{\tau})\).
Subsequently the intervention is  removed at a random system
state, after  which the  distribution of system  states will
gradually converge  back to the original  equilibrium. Nodes
with  higher   dynamic  importance   will  cause   a  larger
difference in the system state probability distribution over
time. Consequently, we quantify the  causal impact of a node
by integrating over time \(t\)  the difference in system state
distribution from  the moment  the intervention  is released
(\(t  = \tau\)).  Since our  model  is discrete  in time,  the
integral  becomes a  summation. Thus,  we define  the causal
impact of node \(s_i \in S\) as

\begin{equation}
\label{eq:causal_impact}
\begin{split}
\Gamma(s_i) &= \sum^\infty_{t= \tau} \gamma(s_i^{t }) \Delta t\\
&= \sum_{t = \tau}^\infty \sum_{s_j \in S}  D_{KL}(p'_{s_i}(s_j^{t}) || p(s_j^{t}) ) \Delta t
\end{split}
\end{equation}

where  \(D_{KL}\) is  the KL-divergence,  and \(\Delta  t =  1\)
throughout  the   paper.  KL-divergence  \(D_{KL}(p   ||  q)\)
quantifies the difference  between probability distributions
and  is  non-negative,   invariant  under  affine  parameter
transformation  and  zero when  \(p  =  q\). For  example,  if
\(\Gamma(s_i) =  0\) the intervention  on nodes \(i\)  caused no
difference in any state probabilities.

The   driver   node  can   then   readily   be  defined   as
\(\textrm{argmax}_{s_i  \in   S}(\Gamma(s_i))\).  A  numerical
implementation  of driver  node identification  is given  in
section \ref{sec:org456a38b}.

In this study  we allow the intervention to  evolve for some
time period  \(\Delta t_{\textrm{nudge}} = \tau  - t_0\) after
which the nudge  is removed (see \ref{sec:org456a38b}  and fig. \ref{fig:methods2}  C). The
intervention  will  transiently  bring  the  system  out  of
equilibrium.  For   ergodic  systems,  the  effect   of  the
intervention will be lost from the system over time. Namely,
\(P_{s_i}'(S),   \forall  s_i\)   will  tend   to  equilibrium
distribution \(P(S)\) from \(\tau\)  on wards. The causal impact
is computed  relative to this \(\tau\)  (fig. \ref{fig:methods2}
C).  As  a  consequence  \(\Gamma(s_i)\) will  be  finite  for
systems under  study here,  but may diverge  for non-ergodic
systems. The  duration \(\tau\) needs to  be set appropriately
that the  intervention is  allowed to percolate  through the
system.  Here  we  used  \(\tau =  15\)  for  random  networks
consisting of  10 nodes and  \(\tau = 25\) for  the real-world
weighted network consisting of 12 nodes (see section \ref{sec:org2760b7b}).

The time  prior to \(\tau\)  is used for computing  the causal
impact   on   a   node   as  the   intervention   could   be
disproportional  affected by  the act  of changing  the node
distribution.  It does  not  accurately  reflect the  causal
impact of the node on the rest of the system. Only the decay
after \(\tau\)  would be proportional  to the causal  impact a
node has on all other  nodes. The causal impact is therefore
computed based on the decay of the causal impact relative to
\(\tau\).

\subsubsection{Intervention size}
\label{sec:org4968a04}
\begin{figure}[htbp]
\centering
\includegraphics[width=.9\linewidth]{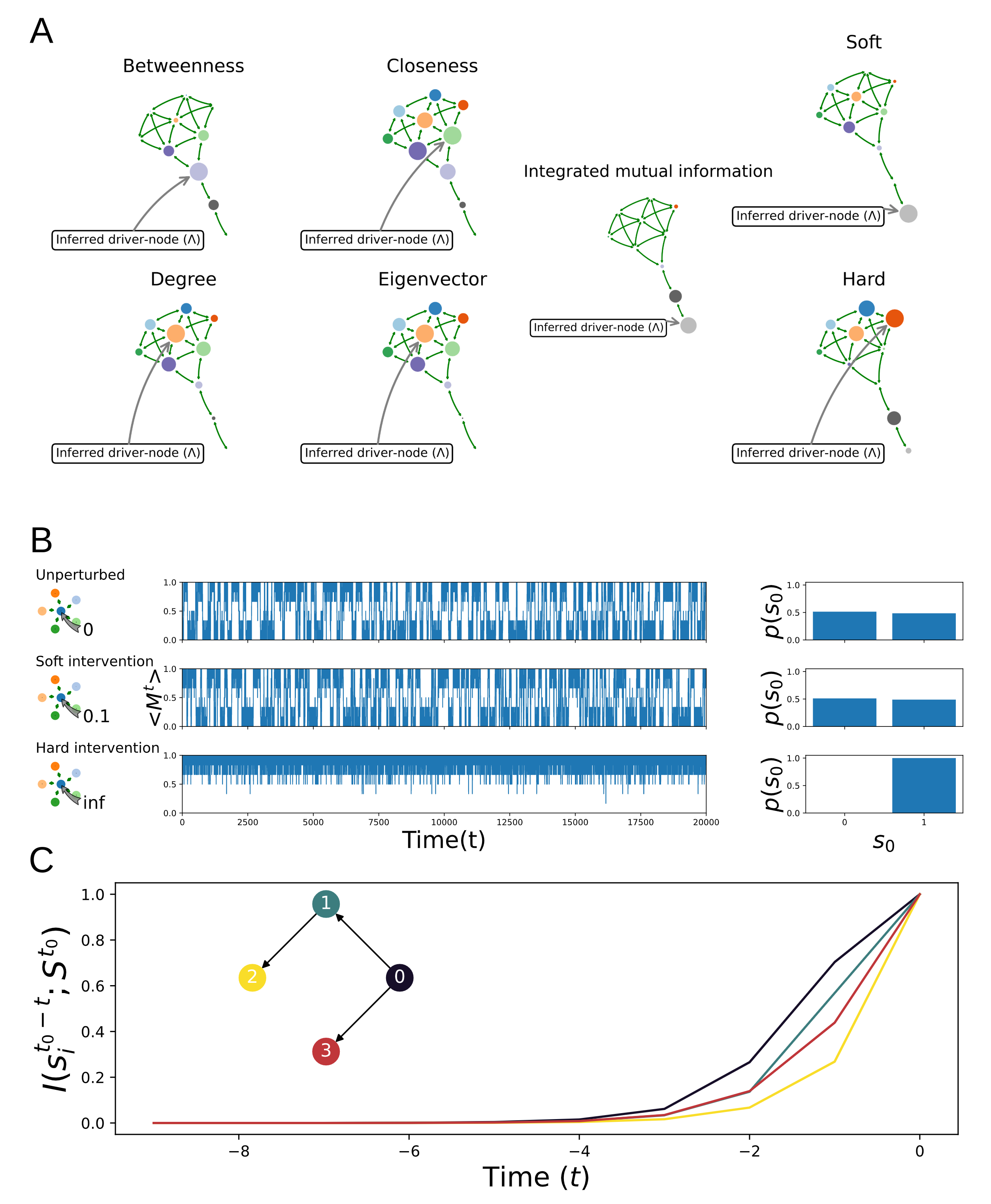}
\caption{\label{fig:methods1}(A) Driver node inference. Dynamics are simulated using kinetic Ising spin dynamics. Causal interventions (soft and hard) are shown in the left column. Structural metrics (betweenness, closeness, degree, eigevenvector centrality) each produce different driver node estimates. Integrated mutual information predicts the driver node for soft causal interventions. High causal interventions produce different system dynamics (see B for an example) and different driver nodes. This figure show that dynamics interact with structure to produce non-trivial driver node estimates. In addition, causal intervention size (hard or soft) influence the observed system dynamics. (B) Effect  of intervention  size on system magnetization \(<M^t> = \frac{1}{n} \sum_i s_i^t\) with kinetic Ising spin dynamics (see \ref{sec:org254cd9f}). Energy is added (gray arrow) to the nodal Hamiltonian of the blue node (\(s_0\)).  (top) Non-Intervened  system dynamics  are shown with the distribution  of the blue node \(s_0\) (right). (middle)  Soft intervention on node \(s_0\)  keeps similar dynamics  as the non-intervened system  dynamics. (bottom) Hard  interventions  yield  profound   different  system  dynamics  compared  to non-intervened dynamics. (C) Example of non-causal inflation of mutual information decay. The network structure is given by the graph inset; consisting of a system with 4 nodes in which each node is binary variable. Node 0 has a 50/50 distribution and all other nodes copy the state of its predecessor. Node 0 has the largest causal influence as it can influence the state of all other nodes in the system (over time). The information content of node 3 is biased; it stores the information from 0, but has no downstream causal effects. Yet, its information decay is similar to 1 which has one downstream node. The true driver node (0) has the largest information decay over time and is not confounded.}
\end{figure}

In experiments concerned  with measuring causal flows in  networks, often ``hard''
causal  interventions  are  used  to   determine  the  causal  impact  of  nodes
\cite{Zhang2017,Liu2016a,Gates2016,Yan2017}. Hard  causal interventions  are those
that  (effectively)   remove  all  inputs   from  a  node.  In   contrast,  soft
interventions,  keep the  existing causal  inputs intact  but add  an additional
causal effect to  the dynamics of a  variable. In general, the  larger the added
causal  effect  of  the  soft  intervention, the  more  this  intervention  will
`overpower' the existing causal effects and hence the more the soft intervention
converges  to  a   hard  intervention.  In  non-linear   dynamical  systems  the
intervention  size is  crucial, since  even small  interventions can  have large
effects, especially  in the presence  of bifurcations. In order  to characterize
the system  without intervention  as much  as possible, one  should thus  aim to
intervene  with a  minimal (but  still  measurable) effect.  To illustrate  this
further, consider  the system  depicted in fig.  \ref{fig:methods1} B,  where each
node  is update  with  eq. \eqref{eq:hastings}.  The figure  shows  the effect  of
intervention  size  on  the  observed  system  dynamics.  Namely,  when  a  hard
intervention  is  used (bottom)  the  system  magnetizes.  In contrast,  in  the
non-intervened  system, the  system magnetizes  periodically, i.e.  there system
dynamics evolve  with time periods  of magnetization and metastable  switches to
the other side of the  magnetization. Contrasting the non-intervened system (top
plot)  with  the bottom  plot  (hard)  intervention,  shows how  the  metastable
behavior disappears. Soft interventions  (middle plot) maintain the meta-stable
behavior of the system.

There will always be a minimal intervention size for which there is a measurable
(i.e. non-zero) causal effect  (eq. \eqref{eq:causal_impact}), given finite amount
of  data. The  causal impact  will  be maximal  for hard  interventions. As  the
intervention size is decreased so will  causal impact. We hypothesize that there
exists  a lower  bound for  the soft  interventions for  which there  will be  a
measurable causal  impact. This  minimal intervention size  is dependent  on the
systems structure  as well  as the dynamics  of the system  and is  difficult to
determine \emph{a  priori}. Yet it is  important to approach this  minimum because if
the intervention  size is too  large, then  the intervened system  dynamics will
diverge  from  the non-intervened  system  dynamics,  losing its  representative
capacity. Therefore, we estimate this  minimum numerically, described further in
section \ref{sec:org81ef225}.

\subsubsection{Intervention in kinetic Ising model}
\label{sec:org94c855d}

Nudge interventions  are implemented  by modifying  the system  Hamiltonian (eq.
\eqref{eq:energy})

\begin{equation}
\label{eq:nudge_implementation}
\mathcal{H}_{\textrm{nudge } s_k}(S) =  -\sum_{i,j} J_{ij} s_{i} s_{j} - \sum_{i} h_{i} s_{i} - \eta s_i \delta_{ik}.
\end{equation}

That is,  an energy term  is added  to the intervened  node \(s_k\) by  letting it
interact with an external `spin state'  \(\eta\). Both soft and hard interventions
were  used  (see section  \ref{sec:org4968a04}).  For hard  interventions  (\(\eta
\rightarrow \infty\))  the nodal dynamics of  the nudge node \(s_k\)  are no longer
influenced by  nearest neighbor  interaction. That  is, the  node state  for the
intervened  node  \(s_j\)  will  not  change  as a  function  of  time.  For  soft
interventions a grid-search  was used with \(\vec{\eta} := \{\eta:  \eta = 0.55 +
i, i\in \{1, \dots, 10 \} \}\) to determine a minimal yet measurable intervention
strength. Once determined the  same \(\eta\) was applied to each  node in turn and
this process is repeated for all networks in the experiments.

\subsection{Measuring information flow}
\label{sec:org1ac74b5}
Each node in a dynamical system can be considered as an information storage unit
\cite{Quax2016a,Quax2017}. For example in social networks gossip can be considered
as information  one person possesses. Similarly,  disease can be present  in one
city  while  being  absent  in  another. Over  time  through  interaction,  this
information stored in  a node will percolate throughout the  system while at the
same time decaying due  to noise. The longer the information of  a node stays in
the system, the  longer it could affect the system  dynamics. Therefore, dynamic
impact of  a node is  upper-bounded by the amount  of information a  node shares
with the entire system \cite{Quax2013,Quax2013a,Quax2017}.

How does  one measure  information stored in  a node? A  node \(s_i\)  dictated by
stochastic and ergodic dynamics can be  considered a random variable. In Shannon
information  theory information  is quantified  in bits,  i.e. yes/no  questions
concerning the  outcome of  a random  variable. The  average information  that a
random variable can encode is called entropy and is defined as:

\begin{equation}
\begin{split}
H(s_{i}) = - \sum_{s_{i} = x} p(x) \log p(x).
\end{split}
\end{equation}

Note all \(\log\) are base 2 in this paper unless specified otherwise.

Entropy  can also  be  interpreted as  the  amount of  uncertainty  of a  random
variable.  In the  extremes the  random variable  either conveys  no uncertainty
(i.e. a node always  assumes the same state), or is  randomly chosen between all
possible states  (uniform distribution). For  example consider a coin  flip. One
may ask  how much information  does a  single coin flip  encode? If the  coin is
fair, i.e. there is  equal probability of the outcome being  heads or tails, the
amount  of questions  needed to  determine the  outcome is  exactly 1.  In other
words, a  fair coin  encodes 1  bit of  information. However,  when the  coin is
unfair the information encoded  is less than one. In the  extreme case where the
coin always turns up heads, the entropy is exactly 0.

The information shared between  a node state \(s_{i}\) and a  system state \(S\) can
be            quantified             by            mutual            information
\cite{Quax2013,Cover2005,Quax2013a,Quax2017,James2017}. Mutual  information can be
informally thought  of as a  non-linear correlation function which  inherits its
properties from  the Kullback-Leibler  divergence. Formally,  mutual information
quantifies the  reduction in uncertainty of  random variable \(X\) by  knowing the
outcome of random variable \(Y\) \cite{Cover2005}:

\begin{equation}
\label{eq:mi}
\begin{split}
    I( X : Y )  &= \sum_{x \in X, y \in Y} p(x, y) \log \frac{p(x, y)}{p(x) p(y)} \\
                &= H(X) - H(X \vert Y),
\end{split}
\end{equation}

where  \(p(x)\)  and  \(p(y)\)  are  the  marginals of  \(p(x,y)\)  over  \(X\)  and  \(Y\)
respectively,  and  \(H(X \vert  Y)\)  is  the  conditional  entropy of  \(X\).  The
conditional entropy \(H(X \vert Y)\) is  similar to the entropy; it quantifies the
reduction  in uncertainty  of the  outcome \(X\)  by knowing  the outcome  of \(Y\).
Please note that  the yes/no question interpretation even  applies to continuous
variables; although it may take an infinite amount of questions to determine the
outcome of a continuous random variable.

\subsubsection{Mutual information and causality}
\label{sec:org92d63a2}
Consider an  Markovian, isolated system consisting  of two nodes  \(S = \{s_i,  s_j\}\) where
\(s_j\) causally depends on the previous  state of \(s_i\), encoded by a conditional
probability distribution. We show here that  the causal impact of \(s_i\) on \(s_j\)
then reduces to mutual information.

For Markovian systems, the future state of the system is independent of its past
given its present  (eq. \eqref{eq:markov}). Therefore, we can  write
\begin{equation}
\label{eq:nearest_neighbor}
\begin{split}
p(s_j^{t+1} | S^t) = p(s_j^{t+1} | s_i^t).
\end{split}
\end{equation}

For any  two nodes  \(s_i, s_j  \in S\),  the KL-divergence  \(D_{KL}(p(s_j^{t+1} |
s_i^t) || p(s_j^t))\) reduces to \(I(s_i^{t+1} ; s_j^t)\).

\begin{theorem}
$E_{s_i} [D_{KL}(p(s_j^{t + 1} | s_i^t) || p(s_j^t)] = I (s_i^{t + 1} ; s_j^t)$ when $s_i$ and $s_j$ have no common neighbors.
\end{theorem}
\textbf{Proof}:

\begin{equation}
\label{app:mi and causal influence}
\begin{split}
		E_{s_i^t}[D_{KL}(p(s_j^{t+1} | s_i^t) || p(s_j^t)] &= E_{s_i^t} \left[ E_{s_j^{t+1} \mid s_i^t} \left[ \log \frac{ p(s_j^{t+1} \mid s_i^t)} { p(s_j^{t+1}) } \right]\right] \\
		&= \sum_{s_i^t} p(s_i^t) \sum_{s_j^{t+1}} p(s_j^{t+1} \mid s_i^t) \log \frac{p(s_j^{t+1} \mid s_i^t)} { p(s_j^{t+1}) }\\
		&= \sum_{s_i^t} p(s_i^t) \sum_{s_j} \log p(s_j^{t+1} \mid s_i^{t}) \log p(s_j^{t+1} \mid s_i^t) - \sum_{s_j^{t+1}} p(s_j^{t+1}) \log p(s_j^{t+1})\\
		&= H(s_j^{t+1}) - H(s_j^{t+1} \mid s_i^t) \\
		&= I(s_j^{t+1} : s_i^t) \blacksquare.
\end{split}
\end{equation}

In this study, the  entire system is considered as a  downstream ``node''. That is
\(I(S^{t+1} ;  s_i^t)\) is  the causal  information flow  since does  there cannot
exist any  confounding node, i.e.  we consider  systems where all  variables are
observed.
\subsubsection{Integrated  mutual information}
\label{sec:org00567e6}
In a network of nodes each causal  relation (edge) is obviously not isolated, so
confounding variables  exist. Therefore, the  mutual information between  a node
state  and a  future state  \(I(S^{t_0 -  t}; s_i^{t_0})\)  cannot be  interpreted
purely causal in general. Namely, this  mutual information could in principle be
created  purely  by  another  variable \(s_j^{t_0  -  \omega}\)  influencing  both
\(s_i^{t_0}\) and  \(S^{t_0 - t}\), even  if there exists no  causal influence from
\(s^{t_0}\) to any other variable.

Assuming the  network itself  is isolated,  the only node  for which  the mutual
information with a  future system state could  not have been fully  created by a
confounding  variable is  the driver  node.  That is,  the driver  node has  the
largest mutual  information with  the future  system state.  If this  were fully
induced by a confounding variable, then a different node would have to have even
larger mutual information  with the same future system state.  In addition, this
must hold for  all future system states.  By definition of the  driver node, this
cannot be true under the condition that the system isolated.

This point is illustrated in \ref{fig:methods1}  C. Note that for all other nodes,
however, it is possible  for its mutual information value to  be inflated due to
non-causal  correlations.  This may  result  to  a non-zero  mutual  information
\(I(s_{i}^{t_{0} - t}  : s_j^{t})\)  among the  two variables  even if  they do  not
depend on each other in  causal manner (fig. \ref{fig:methods1}). Consequently, we
define the integrated mutual information (IMI) for node \(s_i\) as

where \(S^{t_0}\) is the  system state at some time \(t_0\)  and \(s_{i}^{t_{0} - t}\)
is the  state of a node  \(t\) away from  that system state.  At time \(t =  0\) the
value equals \(I(s_i^{t_0};  S^{t_0}) = H(s_i^t)\) for any  node.

Here, undirected networks are considered. Due to detailed balance for undirected
networks there exists a time symmetry  in terms of variable dynamics. This means
that  for the  systems  considered in  this  study \(I(s_i^{t_0  -  t}; S^t_0)  =
I(s_i^{t_0 + t}  ; S^t_0)\) (see appendix \ref{sec:org9e950ef}).
It is computationally  easier to compute \(I(s_i^{t_0 + t}  ; S^t_0)\) rather than
the reverse (see  section. \ref{sec:dist_est}). For directed  graphs, however, the
meaning and interpretation of integrated mutual information changes depending on
the direction in time it is computed (see \textbf{appendix} \ref{sec:org9e950ef}). This effect is outside the scope of the present study and will be the
subject of  future studies. For  undirected graphs,  the causal impact  are time
invariant and is equal forward and backward in time.

\begin{equation}
\label{eq:information_impact}
\begin{split}
\mu(s_{i}) = \sum^{\infty}_{t=t_{0}} I(s_{i}^{t_{0} - t}; S^{ t_{0} }) \Delta t,
\end{split}
\end{equation}

For all ergodic  Markovian systems the delayed mutual  information \(I(s_i^{t_0 -
t}  ;  S^{t_0})\)  will  \emph{always}  decay   to  zero  as  \(t  \rightarrow  \infty\)
\cite{Quax2013,Cover2005}.  This  decay  is  monotonic,  which  follows  from  the
data-processing inequality \cite{Cover2005} and  states that information can never
increase  in Markov  chains without  external information  injection (\textbf{appendix}
\ref{sec:org6d2b506}). The  question is \emph{how fast this  decay takes place
for   each   node}   (fig.   \ref{fig:methods1}  C),   and   consequently   how   much
\emph{informational impact} the node will have on the system.

Taking the entire system as a downstream `node', \(I(S^{t+1} ; s_i^t)\) represents
pure  causal information  flow for  the  driver node  as  long as  there are  no
confounding  variables. For  the rest  of the  study we  will therefore  use the
notation \(I(s_i^{t_0 + t} ; S^{t_0})\).

\section{Methods and network data}
\label{sec:org4457d3a}
\subsection{Network data}
\label{sec:org2760b7b}
\subsubsection{Random networks}
\label{sec:org8ee0c67}
In total  16 Erdös-Rényi  random networks  were generated  consisting of  10 nodes
each. Each  random network  is generated  by first  drawing a  random connection
probability uniformly  from [0,1],  and then  creating each  possible undirected
edge with probability  \(r\). Out of these  16 networks, \textasciitilde{}82 percent  had a single
connected component (fig. \ref{fig:methods2}). Each edge had unitary weight.

\subsubsection{Real-world network: psychosymptoms}
\label{sec:orga34e5dc}

In  addition to  the generated  random  networks, we  also test  a small  weighted
network inferred from  real data. This network differs from  the random networks
in  that  it is  weighted  and  reflecting  interactions  among variables  as  a
consequence of a real-world process, as well as reflecting inferred interactions
from real  data. The network  data originates from  the Changing Lives  of Older
Couples (CLOC) and compared depressive  symptoms assessed via the 11-item Center
for Epidemologic  Studies Depression  Scale (CES-D) among  those who  lost their
partner (N=241) with still-married control group (N=274) \cite{Fried2015}. Each of
the CES-D items were  binarized with the aid of a  causal search algorithm using
Ising  model developed  by \cite{VanBorkulo2014}  and represented  as a  node with
weighted connections (fig. \ref{fig:psycho} D). For more info on the procedure see
\cite{VanBorkulo2014,Epskamp2017,Fried2015}. The  11 CES-D items  are (abbreviated
names  used in  the remainder  of  this text  in brackets):  `I felt  depressed'
(depr), `I  felt that everything  I did was an  effort' (effort), `My  sleep was
restless' (sleep),  `I was  happy' (happy)`, `I  felt lonely'  (lonely), `People
were  unfriendly' (unfr),  `I  enjoyed  life' (enjoy),  `My  appetite was  poor'
(appet), `I felt sad' (sad), `I felt  that people disliked me' (dislike), and `I
could not get going' (getgo).

\subsection{Numerical methods}
\label{sec:org81ef225}
\subsubsection{Magnetization matching}
\label{sec:org4510065}
\begin{figure}[htbp]
\centering
\includegraphics[width=.9\linewidth]{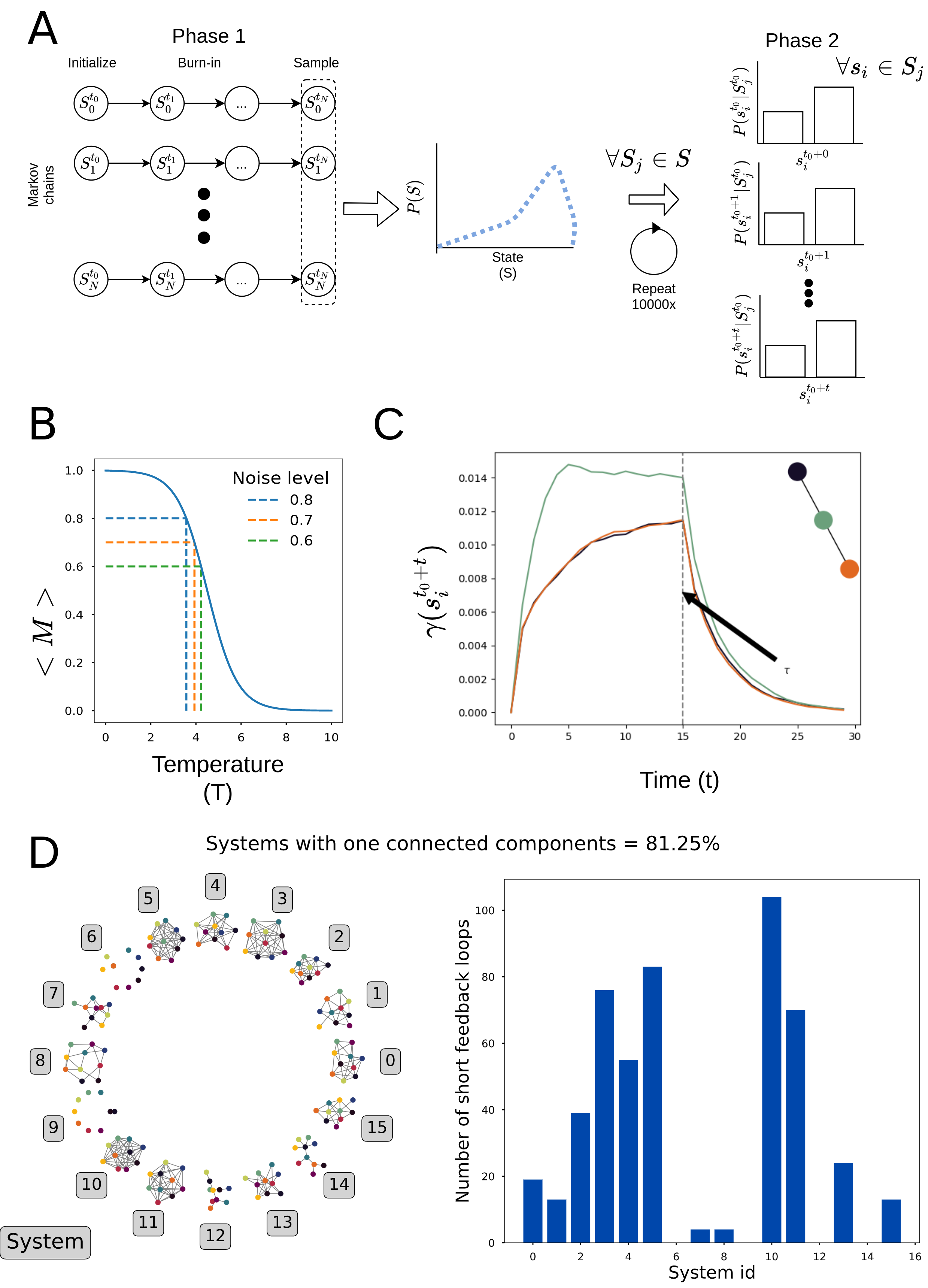}
\caption{\label{fig:methods2}(A) Phase 1 initializes 10000 independent random Markov chains and equilibrated the chains for \(t_n = 1000\) time steps. The state distribution is estimated at \(t_n=1000\) over the 10000  chains. In phase 2 for system states conditional distributions are estimated for \(t\) time-steps from which mutual information is estimated \(I(s_i^{t_0 + t} ; S^{t_0})\). (B) Illustration of temperature matching. The graph depicts the conceptual transition between an ordered state (aligned spins) to a disorered state for the kinetic Ising on complex networks. The noise level was matched to the magnetization ratio of the max magnetization. For increase in temperature the noise level increases. (C) Illustration of applied interventions in an undirected 3 state system (top right corner). Each node is nudged according to eq. \eqref{eq:causal_impact}. The time prior to \(\tau\) ``pushes'' the system dynamics out of equilibrium. From \(\tau\) onwards the causal impact decays proportional to the causal impact a node has on the rest of the system. (D) Generated Erdös-Rényi networks. (left) Structure of each network, the number indicates the system id. (right)Number of triangles (feedback loops) for each system.}
\end{figure}

A prominent feature of the (kinetic) Ising model is the phase change that occurs
as a  function of  noise (fig.\ref{fig:methods2}B)\cite{Glauber1963}.  In this  paper we
tested whether  the amount  of noise  would (i) affect  which nodes  becomes the
driver node  in the system,  and (ii) whether the  correct driver node  could be
predicted using either IMI or network centrality metrics. We tested three levels
of  noise (fig.  \ref{fig:methods2} B);  a  low noise  level (80\%  of the  maximum
magnetization), a  medium noise level (70\%),  and a high noise  level (60\%) were
used. This magnetization  matching was achieved by  estimating the magnetization
curve as a function of \(\beta^{-1}\) numerically.

\subsubsection{Estimating \(p(S^{t_0})\) and \(p(S^{t_0 + t} | S^{t_0})\)}
\label{sec:dist_est}
For  each noise  level,  \(N=\nSamples\)  independent Markov  chains  are run  for
simulation with 1000 steps (fig. \ref{fig:methods2} A). Each chain was initialized
with random  state distribution  over nodes. Each  simulation step  executes the
following:

\begin{enumerate}
\item Pick a node at random from the system with equal probability;
\item Compute energy using eq. \eqref{eq:energy};
\item Flip the node state with probability eq. \eqref{eq:hastings}.
\end{enumerate}

From  this  set,  the  equilibrium distribution  over  states  \(p(S^{t_0})\)  was
constructed in the form of sample of \(N\) system states.

For this  sample of  system states,  Monte-Carlo methods  are similarly  used to
construct the  conditional \(p(s_i^{t_0  + t}  \vert S^{t_0})\).  For each  of the
sample states \(S_i  \in S^{t_0}\), the procedure above was  repeated 100000 times
for  100 time  steps in  the psychosymptoms  and 30  time steps  for the  random
generated networks. All numerical  experiments were repeated \(n_{\textrm{trials}}
=  20\) times  to  provide confidence  intervals  for the  results  in all  nudge
conditions and temperature settings (noise conditions).

\subsubsection{Time symmetry and mutual information}
\label{sec:orgc5bc078}
Thusfar, the definition of causal impact and IMI is ambiguous to the whether \(t\)
is positive or negative.  Namely, if the node state \(s_i^{t  \pm 1}\) is captured
in forward in time  or backward in time with respect to  some state \(S^{t}\). For
undirected networks  with Ising  spin dynamics there  exists time  symmetry with
respect  to how  causal  influence flows  through the  network  due to  detailed
balance (see  \textbf{appendix} \ref{sec:org9e950ef}).  However, for
directed networks this  is not the case. The results  obtained here are obtained
using forward  simulation in time  only. The detailed balance  condition ensures
that the  results would  be symmetric  when simulating  the system  backwards in
time.

\subsubsection{Area under the curve estimation}
\label{sec:orge84277a}
\begin{figure}[htbp]
\centering
\includegraphics[width=.9\linewidth]{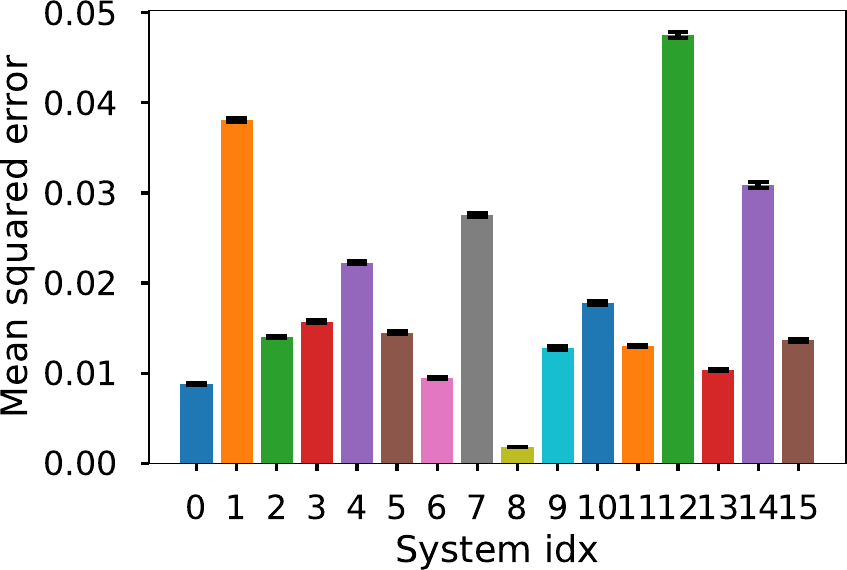}
\caption{\label{fig:fit_error_random}Average mean squared error \(\pm 2 SEM\) per system}
\end{figure}

\begin{figure}[htbp]
\centering
\includegraphics[width=.9\linewidth]{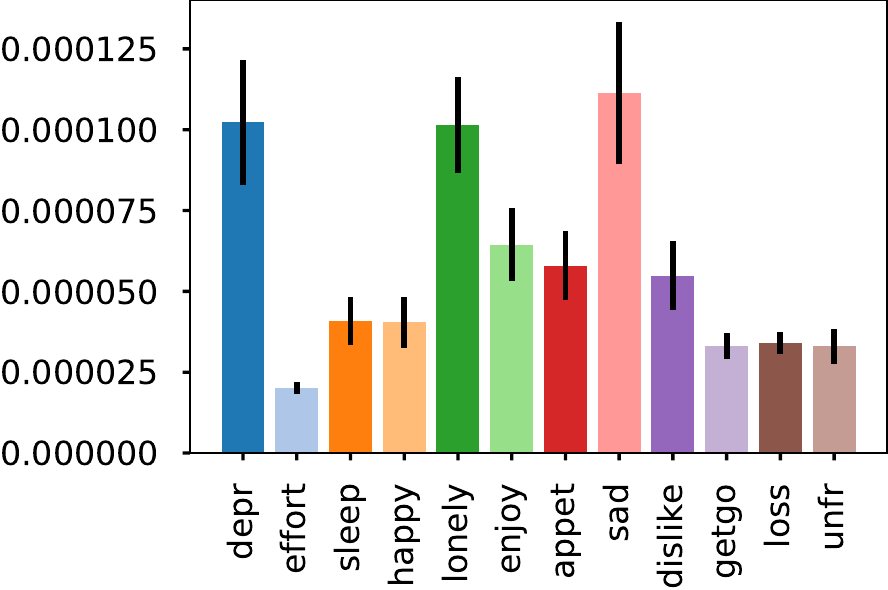}
\caption{\label{fig:fit_error_psycho}Average mean squared error \(\pm 2 SEM\) for psychosymptom system.}
\end{figure}

The mutual information  over time and KL-divergence over time  were scaled for
visual purposes in the range \([0,1]\) per trial set. A double exponential, \(y =
  a \exp(-b (t - c)) + d \exp(- e(t - f) )\), was fitted to estimate these curves
and     subsequently     IMI      (eq.     \eqref{eq:information_impact}     and
\eqref{eq:causal_impact})     using    least     squares    regression     (fig.
\ref{fig:main_results} C,\ref{fig:psycho} C).  The kernel showed to be  a good fit
as   indicated  by   the  low   fit  error   (fig.  \ref{fig:fit_error_random},
\ref{fig:fit_error_psycho}).

\subsubsection{Sampling bias correction}
\label{sec:org5a93a49}
Empirical estimates for mutual information  are inherently contaminated due to
sampling bias.  In order  to correct for  this, Panzeri-Treves  correction was
applied \cite{Panzeri2007}. This  method offer a good performance  in terms of
signal-to-noise and computational complexity.

\subsubsection{Driver node prediction and precision quantification}
\label{sec:org456a38b}
It is  possible for two  or more nodes to  have exactly equal  network structure
features as well  as node dynamics. For example consider  a ring structure where
each node has the exact same connectivity  and all nodes have the same dynamics.
In  this case  each node  must have  the  same causal  effect, and  it would  be
impossible  to disentangle  these  nodes causally  from  one another.  Similarly
graphs that are similar, e.g. show  high degree of structural similarity but are
not isomorphic, this causal separation may proof difficult for finite samples in
stochastic  settings  as was  discussed  in  \ref{sec:dist_est}. Consequently,  we
applied a parametric bootstrap procedure to estimate driver node sets (algorithm
\eqref{alg:driver_set}).

\paragraph{Driver set estimation}
\label{sec:org169f72b}
The  area under  curve values,  e.g.  integrated mutual  information and  causal
impact,  were  resampled  to   generate  bootstrap  distributions  (  \textbf{appendix}
\ref{sec:org603bd1b}). This creates a  confidence interval for the integrated
mutual  information and  causal  impact. From  these  bootstrap distributions  a
driver node set  is estimated (see algorithm  \eqref{alg:driver_set} in \textbf{appendix}
\ref{sec:org603bd1b}). The  bootstrap procedure  constructs driver  node set
\(\Lambda\) iteratively by comparing the overlap  \(\phi\) with of the bootstrap for
each  variable with  the  distribution of  the estimated  driver  node. For  all
experiments  \(\phi  = 0.5\).  The  driver  node  distribution  was taken  as  the
bootstrap distribution with the highest mean.  Variables will be included to the
driver node set \(\Lambda\) if the  overlap between its bootstrap distribution \(i\)
and driver node bootstrap distribution \(j\) was  \(\phi_{ij} > 0.5\). In total \(N =
1e4\) bootstrap trial were constructed of size \(M = n_{\textrm{trials}} = 20\); for each of
the trials the average was computed.  For each variables a Gaussian distribution
was estimated over  the \(N\) bootstrap. This distribution was  used for computing
the overlap with the driver node distribution.

The bootstrap procedure  cannot be applied to the network  centrality metrics as
there  exists  only  one  centrality  rank  assignment  per  network  structure.
Therefore, the  driver nodes as  inferred by  maximum centrality metrics  is the
set  of nodes  (\(\Lambda\)) whose  centrality  metric (\(f\))  equals this  maximum
value, i.e.

\begin{equation}
\label{eq:centrality driver node}
\begin{split}
	  \Lambda_{cent} = \textrm{argmax} f_{cent}(x) := \{ x \in S : f_{cent}(s) \leq f(x) \forall s \in
	  S \}
\end{split}
\end{equation}

where \(f\) is the centrality function which  assigns a real value to each node in
the system. For degree centrality,

\begin{equation}
\begin{split}
f_{deg}(a_i) = \sum_j a_{ij}
\end{split}
\end{equation}

where \(a_{ij}\) is the weighted connectivity between node \(i\) and \(j\) in the
adjacency matrix \(A\) of the network. If \(a_{ij}>0\) node \(i\) and \(j\) are
connected. In this study degree, betweenness, closeness or eigenvector
centrality were used. (see section \ref{sec:org9e3ed34} for the formal
definitions for the centrality measures).

\paragraph{Ground truth comparison}
\label{sec:org97a0860}
The ground  truth values are the  driver node estimations for  the causal impact
bootstrap  distribution.  Each  estimator  also  generated  a  driver  node  set
estimation.  That  is  for  integrated mutual  information,  degree  centrality,
closeseness  centrality,  betweeness  centrality,  eigenvector  centrality,  the
bootstrap  distribution  generates an  estimated  driver  set. To  evaluate  the
performance  of  these  estimators,  an  overlap score  was  computed  with  the
ground-truth (causal impact estimation) using the Jaccard similarity metric:

\begin{equation}
\label{eq:overlap_score}
	J = \frac{A \cap B}{A \cup B}.
\end{equation}

A Jaccard score of 1 means perfect  overlap, i.e. the driver node set identified
by causal impact and  predictor set by IMI or one of  the centrality metrics are
identical.  Conversely,  a   Jaccard  score  of  0   means  completely  disjoint
driver-sets.

For every  network, intervention size  and temperature the similarity  metric we
computed the Jaccard score per predictor. Additionally, the relative performance
ratio of the IMI predictor was evaluated by

\begin{equation}
\label{eq:overlap}
\begin{split}
R_{cent} = J_{IMI} - J_{cent}
\end{split}
\end{equation}

where \(J_{cent}\)  is the Jaccard  score of the structural  metrics (betweenness,
closeseness,  betweeness, eigenvector  centrality).  This performance  indicator
falls within  [-1, 1] range: A value of -1  would indicate that  the centrality
metric correctly  identified the driver  node set; a  score of 0  would indicate
equal performance for the driver node identification between a centrality metric
and IMI; a score of 1 would indicate a correct identification of the driver node
for IMI,  but a false identification  of the centrality metric.  The ratios were
bootstrapped (\(N  = \num{100  000}\)) and  tested for  significance at  \(\alpha =
0.01\). The driver node inferred performance  is bound between \((0, \infty)\).

\subsubsection{Software}
\label{sec:org07176ae}
A general toolbox was developed for  analyzing any discrete systems using IMI,
e.g. Susceptible-Infected-Recovered \cite{Matsuda1994},  Random Boolean networks
\cite{Harvey1997}. The core engine is written in 3.07a with python 3.9.4 and offers
C/C++   level  performance\footnote{$\codebase$},   for  more   information  see
\textbf{appendix} \ref{sec:org6e4c208}.

\section{Results}
\label{sec:org52a5001}
\subsection{Random network results}
\label{sec:orgeb95c62}
\begin{figure}[htbp]
\centering
\includegraphics[height=.86\textheight]{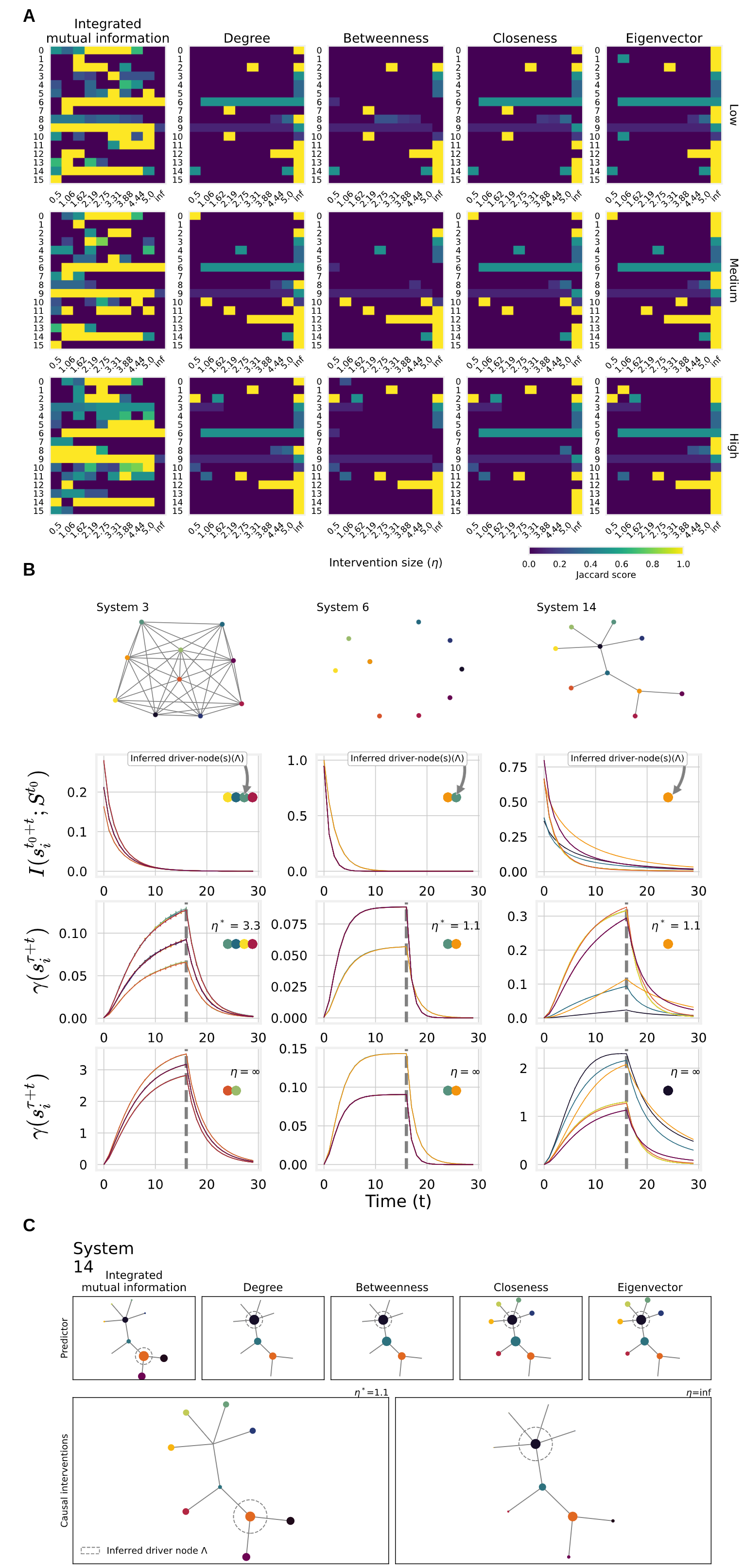}
\caption{\label{fig:main_results}(A) Jaccard score per system (see figure \ref{fig:methods2} D for the various network structures) as a function of intervention size. Each row depicts an increase in noise indicated by the label on the right side of the plot. On average integrated mutual information is a better predictor for the driver node than the centrality metrics used in this study (see also fig. \ref{fig:erdos_stats}). Additionally, centrality metric tend to become predictor for hard interventions (\(\eta = \infty\)), whereas this is not the (generally) so for integrated mutual information. Hard interventions can cause different causal dynamics not present in unperturbed system dynamics (see B/C). Integrated mutual information is based on observations of the system. The Jaccord score indicates that integrated mutual information can infer driver nodes for unperturbed dynamics. (B) Example of typical experimental results. The figures highlight a selection from the generated random networks form A for the low noise condition. (top) The graph structure of the system and the mutual information decay curves (second from top). (second bottom and bottom) Show the causal impact decay for soft intervention \(\eta^*\) and hard intervention \(\eta = \infty\). The gray dotted line indicates \(t = \tau\) where the nudge is removed from the system, the \(t>\tau\) produces causal decay proportional to a node's causal importance. For soft causal interventions the inferred driver node based on integrated mutual information (second from top) is predictive for the true causal driver node (second from bottom) with soft causal interventions. Importantly, the causal driver node may change as a function of the intervention size (second bottom vs bottom plot). The centrality metrics tend to not correctly identify the driver node  (see C for an example). (C) Mutual information decay  (\(I(s_i^{t_0 + t} ; S^{t_0 + t})\)) (top) and causal impact (\(\gamma(s_i^{\tau + t})\) for minimal soft intervention \(\eta^*\) (middle) and hard intervention (bottom).}
\end{figure}

\begin{figure}[htbp]
\centering
\includegraphics[width=.9\linewidth]{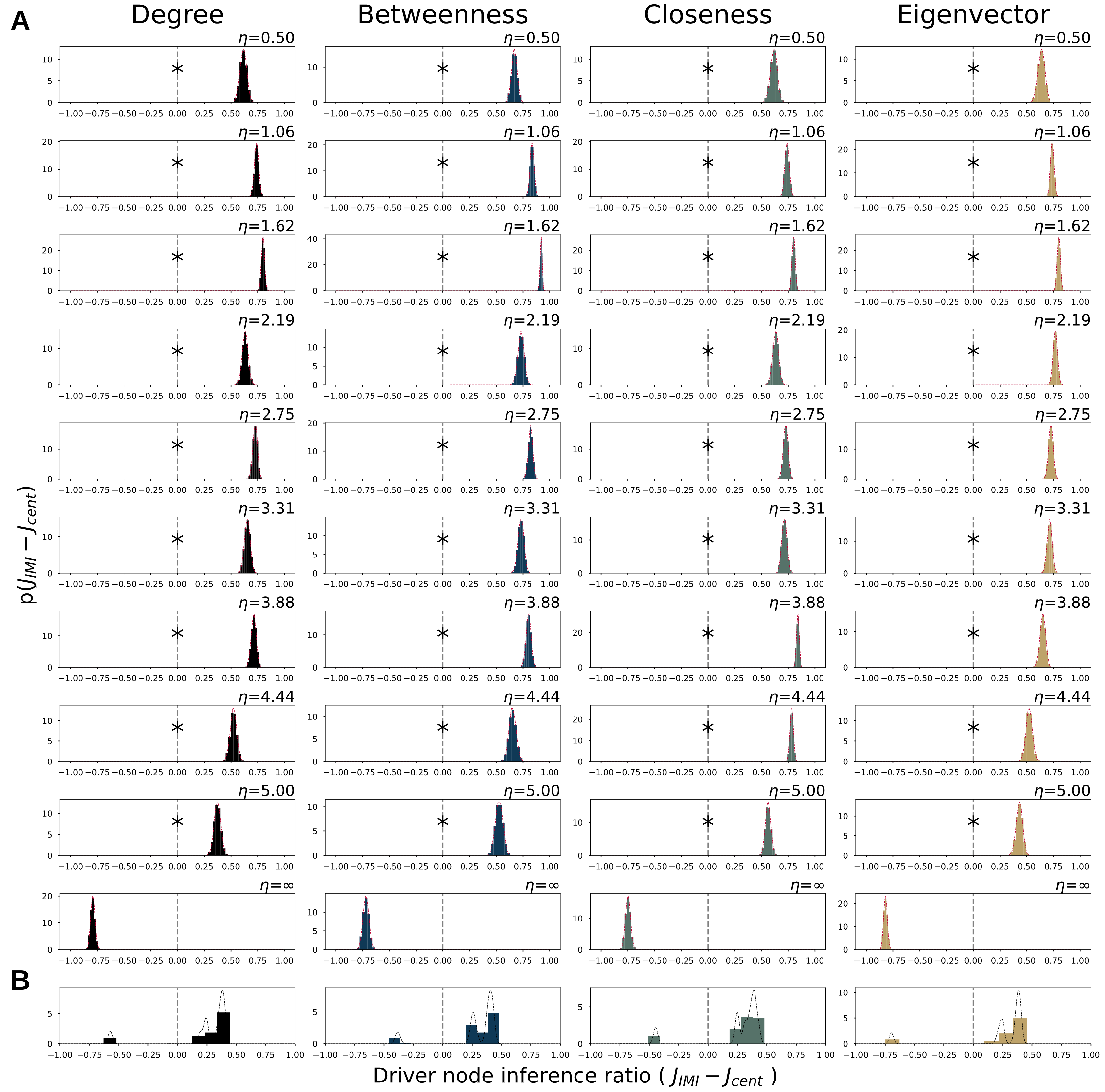}
\caption{\label{fig:erdos_stats}(A) Bootstrap results for driver node inference score \(R_{cent} = J_{IMI} - J_{cent}\) per intervention size and for all interventions and noise  levels(B). A kernel density is fitted for each distribution and integrated over the interval [-1, 0]. An inference ratio of 1 indicates that IMI is better a predicting the driver node than the structural metric and vice versa for a score of 0. Significant intervals are indicated by a asterisk * (\(\alpha = 0.01\)). Except for the hard interventions (\(\eta = \infty\)) intervention size IMI is a better predictor than any of the network structural features.}
\end{figure}

Driver node inference  accuracy is depicted in fig.  \ref{fig:main_results} A for
both IMI and the network centrality measures. Three crucial observations can be
made from  the Jaccard scores.  First, for nearly  all systems there  exists an
intervention size  for which IMI obtains  a Jaccard score of  1 (perfect true driver
node  inference), whereas  this  is  not always  the  case  for the  centrality
metrics, see e.g. system 3, 7, 15.  Secondly, IMI is predictive nearly only for
soft intervention sizes, i.e. intervention sizes smaller or of similar order of
magnitude as the  existing forces acting on nodes(i.e. their  degree in the \(J\)
interaction matrix). In contrast, centrality  metrics are mainly predictive for
hard interventions.  The statistical  results reflect these  observations (fig.
\ref{fig:erdos_stats} A and B).

In figure \ref{fig:main_results}  B average decay curves are shown  for IMI (top)
and different intervention  sizes (middle) and hard  interventions (bottom) for
the  systems   2, 5, 12.  Their   network  structure  is  depicted   in  fig.
\ref{fig:main_results} B (top). Comparing hard interventions (bottom) with a more
moderate intervention  strength (middle)  in system 2,  shows that  the driver
node can significantly differ depending on the intervention size applied to the
system. The  order of  the causal  importance is  noticeably different  for the
minimal  intervention \(\eta^*\)  versus  the  strong intervention  \(\eta=\infty\)
(middle and bottom plot in figure \ref{fig:main_results} B).

\subsection{Real-world network: psychosymptoms results}
\label{sec:orgb166d9d}
\begin{figure}[htbp]
\centering
\includegraphics[width=\textwidth]{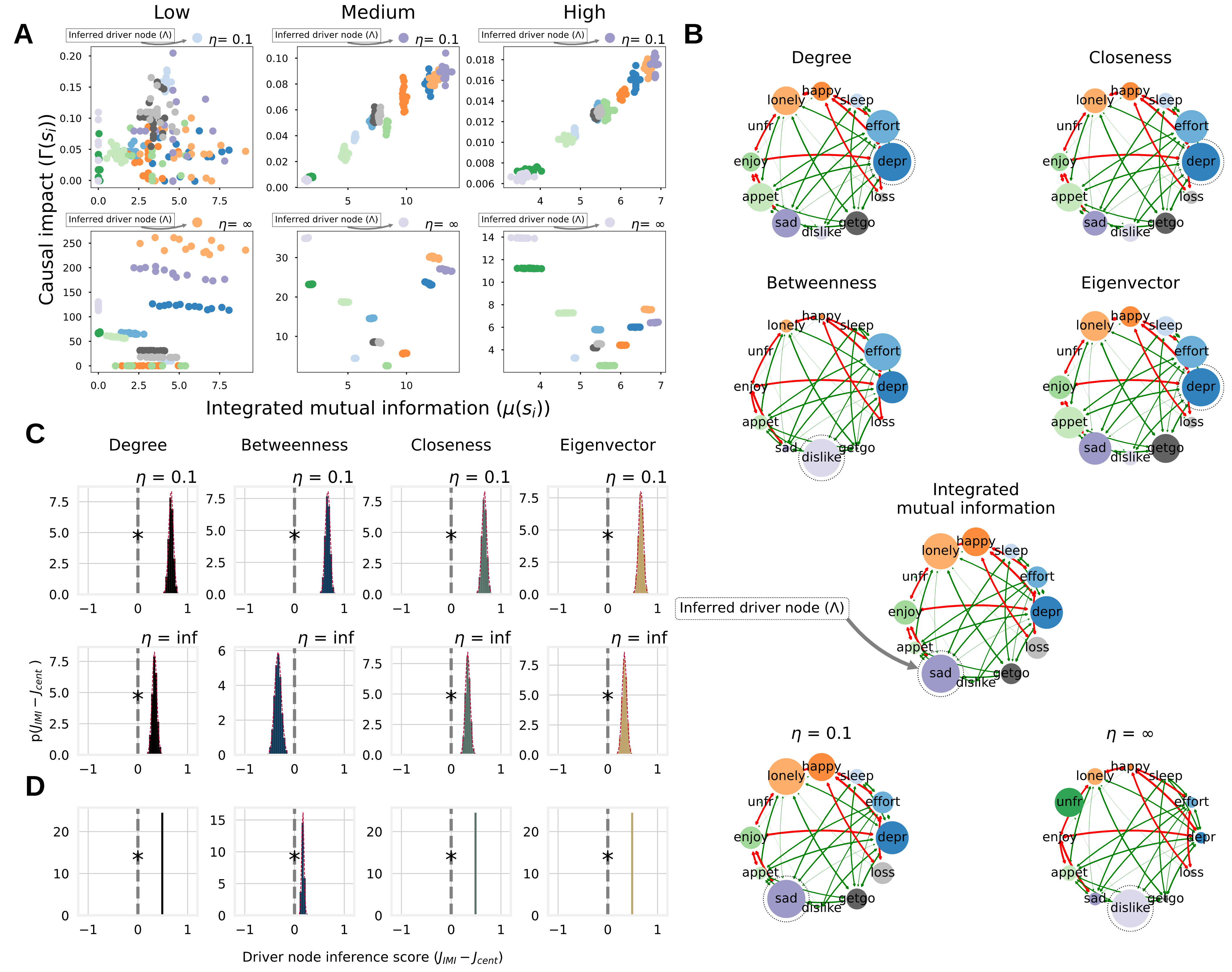}
\caption{\label{fig:psycho}(A) Causal impact versus integrated mutual information for different noise levels and intervention sizes. The inferred driver node is indicated above each figure with a dot. For low noise level the numerical estimates were too noise and driver estimation was inaccurate. For intermediate to high noise level, causal impact scale linearly with integrated mutual information for soft interventions. Causal driver node was wrongly predicted for hard interventions by integrated mutual information. Notice how the causal flows change as a function of intervention size (top versus bottom plot). (B) Driver node identification in real-world network of psychosymptoms for medium noise as a function network features (top 4 subplots) and integrated mutual information (middle), and true causal driver nodes per intervention size (bottom). Integrated mutual information correctly identified the driver node with soft interventions, but not for high interventions. Only betweenness centrality correctly identified the true driver node for high causal interventions. (C) Bootstrap distributions for driver node inference score as function of intervention size across noise levels. A score of 1 indicates that integrated mutual information identified the driver node correctly, but the structural metric didn't, and vice versa for a score of -1. In red kernel density estimates are indicated and integrated over [-1, 0]. The asterisk (*) indicates a significant better inference score for integrated mutual information (\(\alpha < .01\)). Integrated mutual information was a better predictor than any of the network features for soft interventions. (D) Bootstrap distribution for driver node inference score as function of temperature and nudge size. The integrated mutual information was a significant better predictor across noise and intervention size (\(\alpha <0.05\) indicated by *).}
\end{figure}

The psychosymptoms system  reveals similar results to the  random networks (fig.
\ref{fig:psycho}).  Namely,   for  medium   to  high   noise  level   IMI  yielded
significantly  higher Jaccard  scores  than centrality  metrics  for low  causal
intervention  (fig.   \ref{fig:psycho}  C/D,   \(p<<0.01\)),  while  not   for  hard
interventions. In contrast, hard interventions yield different causal structures
altogether  which do  not reflect  non-intervened dynamics  (fig. \ref{fig:psycho}
A/B).  For  example the  true  driver  node  under hard  interventions  identify
`dislike' (medium and  high noise) to be the driver  node (fig. \ref{fig:psycho} A
). Whereas for low causal intervention  `sad' is identified as driver node (fig.
\ref{fig:psycho}  A). This  implies that  intervention itself  has impact  on what
causal  structure  is observed  and  that  the  intervention can  show  systemic
behavior  not present  in the  non-intervened system.  The soft  intervention of
\(\eta =  0.1\) was  too low  to provide proper  resolution for  identification of
driver nodes in the low noise setting (fig. \ref{fig:psycho} A). Hard intervention
in  low noise  condition  did provide  a  different driver  node  than the  soft
intervention.  The grid-search  for optimal  \(\eta^*\) was  insufficient for  the
psychosymptom network and should be investigated in future studies.

In  addition, the  change in  driver node  observed in  figure \ref{fig:psycho}  A
highlights  one major  flaw in  centrality metrics:  they cannot  account for  a
change in driver nodes  due to a change in dynamics.  The implicit assumption on
dynamics  that each  centrality metric  holds,  provides only  one estimate  per
network structure. In contrast, IMI does  not depend on what mechanisms generate
system  dynamics. Instead,  it uses  the distribution  dictated by  these system
dynamics which match the true driver nodes.

Fried  and colleagues postulated  that  `lonely' was  the  gateway from  which
information spreads through the network, i.e. bereavement was embodied mainly by
`loneliness'  which   then  percolated   its  effect   to  the   other  symptoms
\cite{Fried2015}.  Since the  data was  cross-sectional, the  comparison with  the
results  from this  study  relies on  the assumption  that  binary dynamics  are
representative  of  the  absence  and presence  of  psychological  symptoms.  If
correct,  the  results  from  this  study  give  a  causal  perspective  on  the
associative results from \cite{Fried2015}. The  results from this study postulate
that `depr', `lonely' and `sad' have  similar causal effect for moderate to high
thermal noise.

It is important to emphasize that a quantification is given in terms of absolute
effect  size and  not directed  effects. This  means that  nudging for  instance
`sleep' has  some effect \(X\)  on the  psycho-symptom network, in  what direction
that  effect  is, or  whether  it  has a  positive  or  negative effect  on  the
bereavement score /  cognitive load of the  patient is not clear,  and should be
the subject of future studies (see \textbf{appendix} \ref{sec:org16cff9c}).

\section{Discussion}
\label{sec:org6ff7365}
Structural metrics can be  considered as implicitly assuming
a  particular  dynamics \cite{Borgatti2006,Borgatti2005}.  For
example, the  eigenvector centrality has a  clear analytical
connection with linear dynamics,  such as a simple diffusion
process.  Betweenness centrality  on the  other hand  can be
considered  to  assume  that   causation  between  nodes  is
transmitted mainly along the shortest paths between pairs of
nodes  (see  \textbf{appendix}  \ref{sec:org9e3ed34}),  such  as  in
message  or  package  routing. Finally,  many  more  network
centrality  metrics have  been developed  based on  specific
dynamics,  such as  current  and  flow centrality  measures.
Thus, such network structural  metrics could indeed be used,
but a careful analysis of  the dynamic equations is required
to  assess  which  centrality  measure (if  any)  turns  out
appropriate.

The results from this study show that high degree nodes with
Ising spin dynamics tend to  actually \emph{not} be a driver node
for small interventions. In this  model it can be understood
since  high degree  nodes exhibit  ``frozen'' behavior;  their
large(r)  number  of  neighbors  effectively  sum  up  to  a
constant  and  strong  force  towards  the  majority  state.
Consequently,  a   constant  (soft)  intervention   will  be
relatively ineffective  with hubs. For  non-intervened Ising
dynamics,  it makes  intuitive sense  that the  dynamics are
driven by  the nodes which  are neither ``frozen''  (hubs) nor
are poorly connected (low-degree  nodes), which is reflected
by the  soft intervention setting. This  insight is specific
to  the Ising  system but  illustrates that  the information
flow and causal flow of a system is not merely determined by
the  connectedness of  a node  in the  network. Rather,  the
connectedness of the entire system in addition to inter-node
dynamics combined  are crucially important for  causal flows
and driver node identification.

The  results further  imply  that  for unperturbed  dynamics
structural metrics may not be predictive for determining the
causal  importance  of  nodes. Soft  interventions  revealed
different  causal structure  than  hard interventions  (fig.
\ref{fig:main_results}      \ref{fig:psycho}).     For      hard
interventions, structural  metrics do become  predictive for
causal importance  for the  networks studied  here. However,
the dynamics of these systems  are shown to deviate from the
non-intervened  dynamics, causing  system dynamics  that are
not representative of the non-intervened system. This can be
seen in the causal  influence in figure \ref{fig:main_results}
and  \ref{fig:psycho}  where  the causal  influence  for  hard
interventions  is  opposite  to soft  causal  interventions.
Consequently, if the aim is  to provide understanding to the
information  flows for  non-intervened  dynamics, then  hard
interventions are not preferred.

In  addition, the  results imply  that in  order to  achieve
maximum impact  for a fixed `intervention'  budget (injected
energy), choosing  the high degree nodes  is not necessarily
optimal.       The      adjusted       Hamiltonian      (eq.
\eqref{eq:nudge_implementation})  introduces  a  bias  on  low
degree nodes.  Here, a causal intervention  was performed by
adding   fixed  energy   to  the   Hamiltonian.  For   fixed
intervention size \(\eta = c\), the causal impact on the nodal
distribution will  be relatively  higher for nodes  with low
degree than  nodes with  higher degree. Higher  degree nodes
may  have higher  causal  effect in  principle  if the  same
probability  mass  is  changed.  However,  moving  the  same
probability mass  scales non-linearly in kinetic  Ising (eq.
\eqref{eq:energy}). For a limited  `intervention budget' it is
preferred  to  locate  those  elements of  the  system  that
reaches maximal causal effect.

For, some  systems, however, the inferred  driver node never
matched  the true  driver  node(s) (e.g.  system  8 in  fig.
\ref{fig:main_results}).  This  could  be   due  to  two  main
reasons.   First  a   grid-search   was   applied  for   the
intervention  size. It  was  argued that  there  would be  a
minimal  intervention   \(\eta^*\)  which  would  lead   to  a
measurable effect.  It is possible that  the parameter space
used here missed the  intervention size that provided enough
resolution  to  accurately  determined the  driver  node(s).
Second, the  numerical procedure  for driver  node inference
could be optimized.  The overlap of distribution  was set to
\(\phi =  0.5\), to  infer the driver  node set  \(\Lambda\), to
prevent false  positives for driver node  identification due
to noise in  observed system states. However,  the choice of
parameter  was not  optimized  and could  lead to  ambiguous
driver node inference. System 8 in particular was most nodes
in the  network had a  similar connectivity pattern,  and as
such causal isomorphy occurs,  i.e. the causal importance of
a  node is  indistinguishable  from any  other  node in  the
system. The  bootstrap estimates led  to high level  but not
perfect of overlap (see \textbf{appendix} \ref{sec:org603bd1b}).
Consequently, if  \(\phi\) was  set differently,  the inferred
driver nodes could be improved. In future studies, we aim to
further look into what how this numerical procedure could be
improved for  inferring driver  nodes in  dynamical networks
particularly for causal isomorphic nodes.

\subsection{Limitations}
\label{sec:orgd145dbe}
The systems considered are discrete and ergodic. IMI assumes
that   the   data-processing   inequality  holds   for   the
system(section   \ref{sec:org00567e6}).   The
data-processing inequality  in ergodic systems  ensures that
\(I(s_i^{t_0 + t} ;  S^{t_0})\) monotonically approaches zeros
as \(t  \rightarrow \infty\) (see  \textbf{appendix} \ref{sec:org6d2b506}). As a consequence IMI  will always be finite for
ergodic    systems.    For    non-ergodic    systems,    the
data-processing   inequality   can   only   guarantee   that
\(I(s_{i}^{t_{0}  + t}  ;  S^{t_{0}})\) never  increases as  a
function  of  \(t\).  Namely, the  data-processing  inequality
ensures  that  no  local  manipulation  of  information  may
increase the  information content of a  signal. This implies
that as  \(t \rightarrow  \infty\), IMI  may not  converge for
nodes with non-zero baselines.  For these cases, however, it
is possible determine the driver nodes by considering finite
time-scales,  or subtracting  the asymptotic  \(\gamma\) value
and reporting it separately.

In addition, we merely studied on type of dynamic here, i.e.
the kinetic Ising model.  Integrated mutual information does
not  assume a  particular type  of dynamic,  i.e. it  merely
requires  the  data processing  inequality  to  be true.  We
believe  that  the  methods  proposed here  can  readily  be
applied  to other  ergodic systems  with different  kinds of
dynamics and produce reliable  estimates for the driver node
under  theoretical  constrains  used   in  this  study  (see
\ref{sec:org8af9ae7}). However, future work would need to
study these claims in more detail.

Systems of  size at most \(n=12\)  were used. The size  of the
system was chosen  due in order to  provide high reliability
of the probability distributions. In addition, larger graphs
can  be  decomposed  in  various  different  network  motifs
\cite{Alon2007}.  It is  believed that  these motifs  form the
``computational'' buildings blocks  for larger complex systems
under the assumption of  nearest neighbor interactions. That
is, understanding  the motifs would  gain insights in  how a
macroscopic  property emerges  from local  interactions. The
aim of this study was  to relate structural connectedness to
dynamic importance; the exact  nature or occurence of motifs
were not the focal point. In real-world systems, however, it
is exactly  the composition and interaction  of these motifs
that are  vital to  complex systems.  Here, the  motifs were
implicit  on  the  real-world   network  and  the  generated
structures. We  leave it up  to future  work to map  out the
driver nodes of common  network motifs in different dynamics
using and relate the  structural importance to the dynamical
importance.

\section{Conclusions}
\label{sec:org7f1d73e}
Our   results  indicate   that  dynamic   importance  cannot
necessarily  be reliably  inferred  from network  structural
features alone,  demonstrated here using kinetic  Ising spin
dynamics. The goal of this paper was to show that structural
methods can provide unreliable  estimates of the driver node
in dynamical systems. The results  from this study show that
the   common   assumption   of   structurally   central   or
well-connected nodes  being simultaneously  dynamically most
important  is not  necessarily  true. This  implies that  we
cannot abstract away the dynamics of a dynamic system before
inferring driver  nodes. The proposed  information theoretic
metric, integrated mutual information (IMI), was better able
identify  the driver  node  for  non-intervened dynamics  in
systems.  Importantly,  IMI  does not  require  knowing  the
dynamics  equations  and/or  the network  structure  of  the
system.   It  is   instead   calculated   directly  from   a
cross-section   of  time-series   of   the  system   without
interventions.  The  proposed  metric could  potentially  be
useful  in  applications  with  rich  data  sets  and  where
performing interventions are infeasible or impractical.

\section{Acknowledgment}
\label{sec:org3f284ea}
This  research is  supported by  grant Hyperion  2454972 of  the Dutch  National
Police. In  addition, Rick Quax  acknowledges funding from the  European Union’s
Horizon 2020 research  and innovation program under grant  agreement No 848146
(ToAition).

We thank  M.Sc. Fiona  Lippert, Dr. Paul  Duijn and Dr.  Thijs Vis  for fruitful
discussions in support  of this paper. We declare that  no conflict of interests
are present in the conduction of this study.

\printbibliography

\section{Appendix}
\label{sec:org2414765}
\subsection{Data-processing inequality}
\label{sec:org6d2b506}
The data-processing inequality can be used to show that no clever manipulation of
the data  can improve the inferences  made from that data  (see for more detail \cite{Cover2005}).

\begin{Definition}
	Random variables $X\rightarrow Y \rightarrow Z$ are said to form a Markov chain if the conditional distribution of $Z$ depends only on $Y$ and is conditionally independent of $X$. Specifically, $X,Y,Z$ form a Markov chain if the joint probability can be written as:
\end{Definition}

\begin{equation}
\begin{split}
p(x, y, z) = p(x) p(y \vert x) p(z \vert y)
\end{split}
\end{equation}

\begin{theorem}
		(Data-processing inequality) If $X \rightarrow Y \rightarrow Z$, then $I(X; Y) \geq I(X; Z)$.
\end{theorem}

\textbf{Proof:} By the chain rule, the mutual information can be expanded in
two different ways:

\begin{equation}
\label{chain rule}
 \begin{split}
	 I(X; Y; Z) &= I(X ; Z) + I(X; Y \vert Z)\\
				&= I(X; Y)  + I(X; Z \vert Y)
 \end{split}
 \end{equation}

Since \(X\) and  \(Z\) are conditionally independent given \(Y\),  we have \(I(X; Z
	\vert Y) = 0\). Conversely, if \(I(X; Y \vert Z) \geq 0\), this would give

\begin{equation}
\begin{split}
	I(X; Y) \geq I(X; Z).
\end{split}
\end{equation}
Thus we only have  equality if and only if \(I(X; Y \vert  Z) = 0\) for Markov
chains. Similarly, one can prove that \(I(Y; Z) \geq I(X; Z) \blacksquare\).

\begin{Corollary}
	\textrm{If} $Z = g(Y) \rightarrow I(X;  Y) \geq I(X; g(Y))$
\end{Corollary}

\textbf{Proof:}  \(X \rightarrow  Y \rightarrow  g(Y)\) forms  a Markov  chain
\(\blacksquare\).

This result implies that no function \(g(Y)\) can increase the information
about \(X\).

\begin{Corollary}
	\textrm{If} $X \rightarrow Y \rightarrow Z$, \textrm{then} $I(X; Y \vert Z) \leq I(X; Y)$
\end{Corollary}

From eq. (\ref{chain rule}) it is noted that  \(I(X; Z \vert Y) =  0\) due to
the definition  the Markov chain  and \(I(X; Z)  \geq 0\).
Therefore:

\begin{equation}
\begin{split}
	I(X; Y \vert Z) \leq I(X; Y) \blacksquare
\end{split}
\end{equation}

The  dependence of  \(X\) and  \(Y\) is  decreased or  remains unchanged  by the
observation of  a ``downstream'' random  variable \(Z\). For any  complex system
in which  the state distribution follows  a Markov chain, i.e.  \(X^{t_0} \to
	X^{t_0+1} \to \dots \to  X^{t_0+\infty}\). The mutual information \(I(X^{t_0};
	X^{t_0 + t})\) will always decay to zero as \(t  \to \infty\).

\newpage

\subsection{Mutual information and time symmetry}
\label{sec:org9e950ef}
\begin{figure}
	\centering
	\begin{subfigure}{.45 \textwidth}
		\centering
		\includegraphics[width= \textwidth]{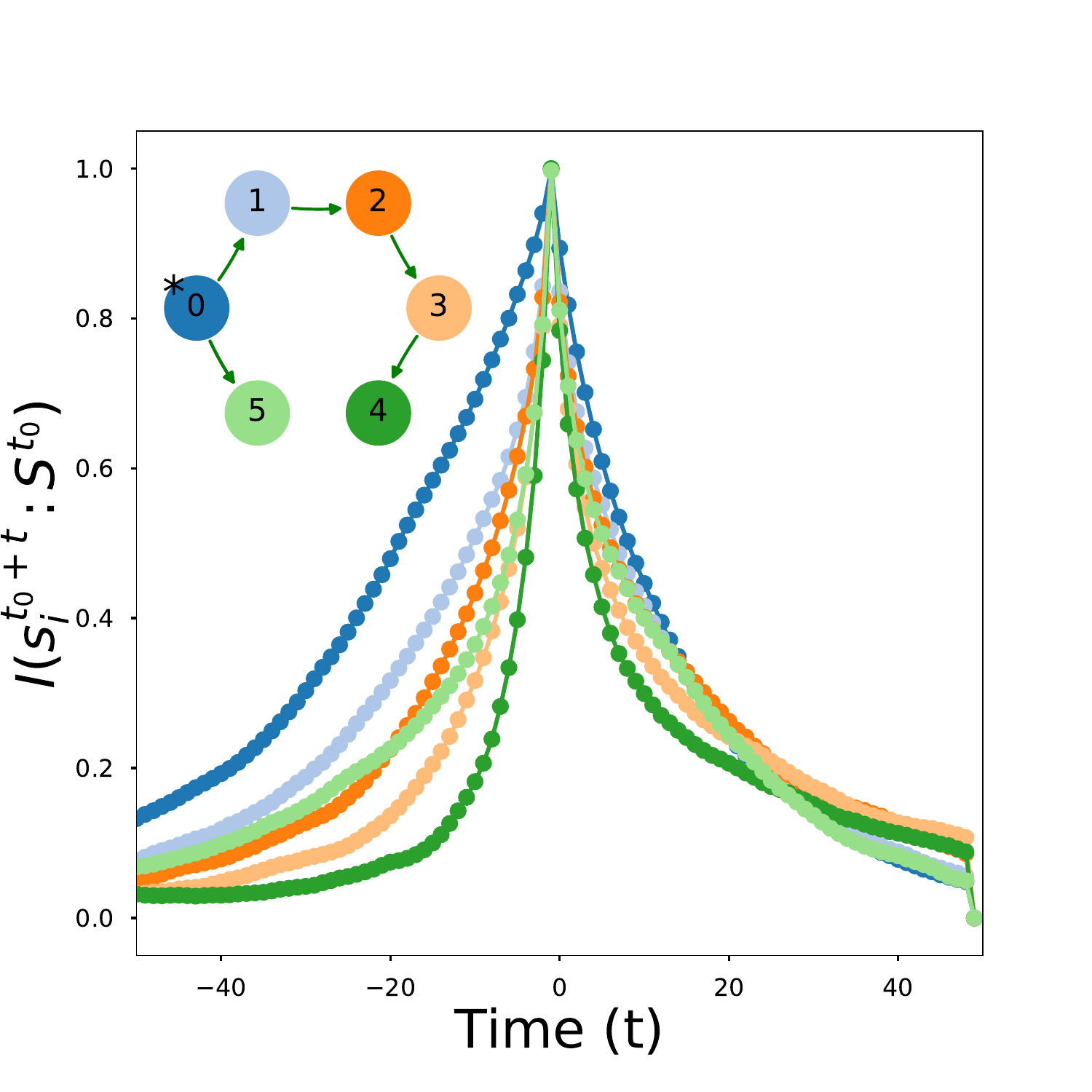}
		\caption{}
		\label{fig:directed asymmetry}
    \end{subfigure}
	~\hspace{0em}%
	\begin{subfigure}{.45\textwidth}
		\centering
		\includegraphics[width= \textwidth]{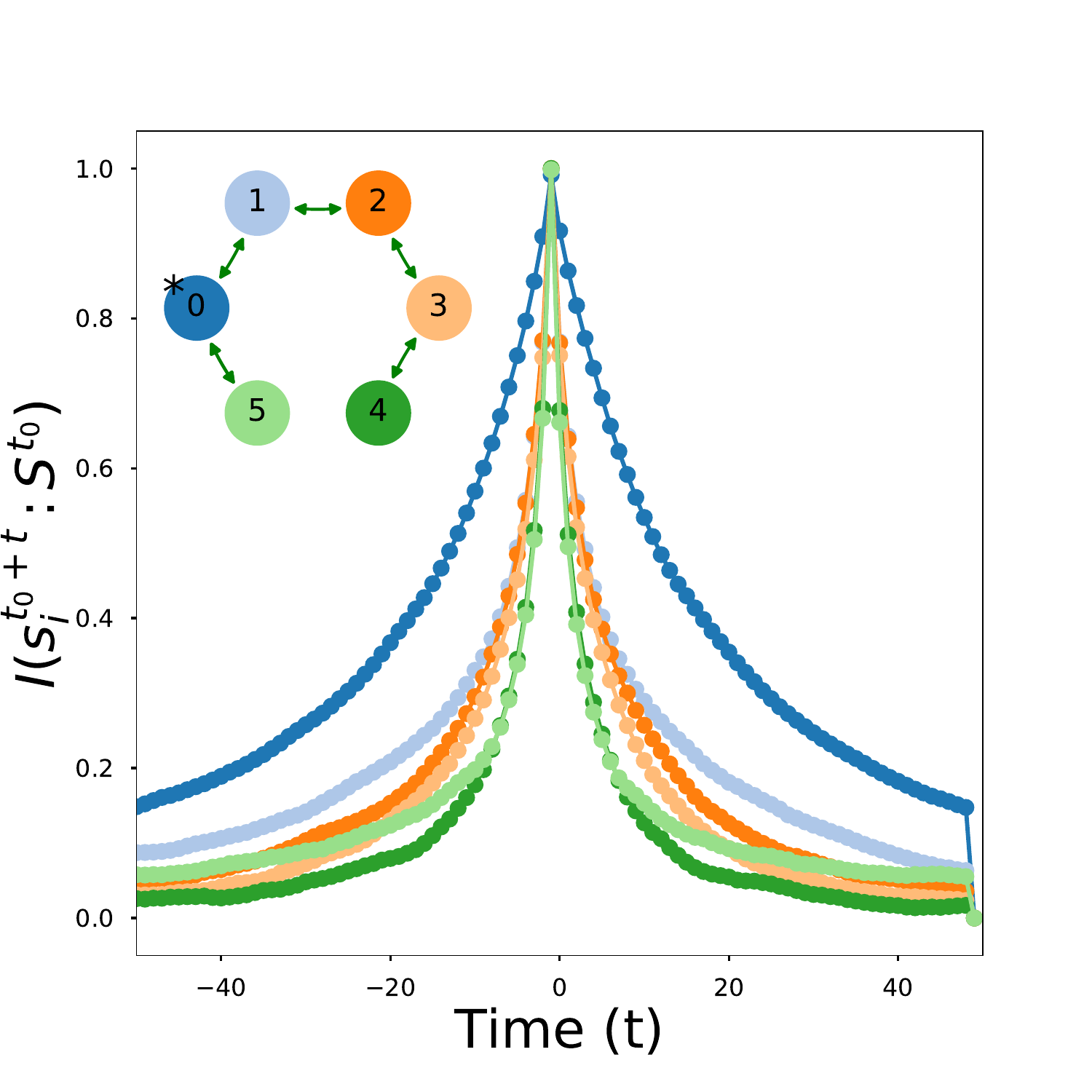}
		\caption{}
		\label{fig:undirected symmetry}
	\end{subfigure}
	\caption{Example of time symmetry in directed and undirected networks. Figure \textbf{\ref{fig:directed asymmetry}} shows the asymmetry that occurs when information flow is directed. The time before the system state $t < t_0$ can be interpreted as information sending. Namely, nodes that have the most impact on the current system state $S^{t_0}$. In contrast, information for $t > t_0$ as information receiving; nodes that receive information from $S^{t_0}$. The most striking example is node 4 which has a sharp decay for $t < t_0$ but a relatively fat tail for $t>t_0$. This change is due to the difference in meaning of the IMI, e.g. sending vs receiving. \newline figure \textbf{\ref{fig:undirected symmetry}} shows that for undirected networks there is no difference between node importance before or after $t_0$; information flows both directions.}
	\label{fig:the symmetry of time}
\end{figure}

The  methods  applied in  the  main  text imply  that  the  metric can  be  used
symetrically. In this  study time-delayed mutual information was  performed in a
`forward' manner for  pratical purposes. Namely, the system  state was simulated
for positive  \(t\) from some \(t_0\).  For undirected networks there  is a symmetry
with  regard to  where  information flows.  Information is  not  bounded by  any
directionality of edges (fig. \ref{fig:undirected symmetry}).

It is important to emphasize that this  (generally) is not the case for directed
networks. If  information is constricted  to flow  in one direction,  the mutual
direction of  time simulation is  crucial. Additionally, directed  networks show
that the metric can be applied for  different purposed. This can be seen in fig.
\ref{fig:directed  asymmetry},  where  forward  simulations  gives  `information
sinks' and backward  simulation provides `information sources'.  IMI in directed
networks will provide information about what nodes receive the most information
over time. In contrast, simulating backwards shows what nodes have most impact on
the instantaneous state of the system.  This dual-use of information will be the
focus of future studies.

\subsubsection{Time reversibility and detailed balance}
\label{sec:orgc6d35fc}
In order to show  that \(I(s_i^{t+1} ;
S^t) = I(s_i^{t-1} ; S^t)\),  we have to show that \(p(S^{t+1}_* | S^{t}_*) = p(S^{t-1}_* | S^{t})\).
\paragraph{Two-sided Markov Chains}
\label{sec:org89fea44}
For  a  positive recurring  Markov  chain  \(\{S^t  :  t \in  \mathbb{N}\}\)  with
transition matrix  \(P\) and stationary distribution  \(\pi\), let \(\{S^t_* :  t \in
\mathbb{N}\}\) a  stationary version of this  chain, i.e. \(S^0 \sim  \pi\). We can
construct  a  two-sided extension  of  \(S^t\)  by defining  a  shift  \(k \ge  1\),
extending the process \(S\) backwards in time: \(S^t_*(k) = S^{t-k}_*\) with \(0 \leq
k < \infty\). It is true that by stationarity \(S^{t-k}_*\) has the same stationary
distribution as \(S^t\): we arrive at \(\{S^t_* : t \in \mathbb{Z}\}\).
\paragraph{Detailed balance}
\label{sec:org1281be7}
Let  \(\{S^t_* :  t  \in \mathbb{Z}\}\) be  a two-sided  extension  of a  positive
recurrent Markov  chain with transition  matrix \(P\) and  stationary distribution
\(\pi\). A transition from state \(i\) to \(j\) is denoted with notation \(i \rightarrow j\).

\begin{equation}
\label{}
\begin{split}
p(S^{1}_* = j | S^0 = i)  &=  p(S^{1}_*(k = 1) = j | p(S^{0}_* (k = 1)) = i)\\
&= p(S^{0}_* = j | S^{-1}_*  = i)\\
&= \frac{p(S^{-1}_* = i | S^0_* = j) p(S^0_* =j)}{p(S^{-1}_* = i)}\\
&= \frac{\pi_{j}}{\pi_i}P_{i \rightarrow j}
\end{split}
\end{equation}

In other words the time-reverse Markov  chain is a Markov chain with transition
probabilities:

\begin{equation}
\label{}
\begin{split}
P_{i \rightarrow j}\pi_{i} = \pi_j P_{j \rightarrow i}
\end{split}
\end{equation}

This is also  known as detailed balance. In the  manuscript, ergodic systems are
used and therefore  satisfy the time-symmetry. This results  that \(I(s_i^{t+1} ;
S^t) = I(s_i^{t-1} ; S^t)\).

\subsection{Data correction and fit errors}
\label{sec:org253fb70}

\newpage

\subsection{Validation of psychosymptoms}
\label{sec:org16cff9c}
The  results  from this  study  imply  that for  low  thermal  noise not  enough
resolution was possible to reliably estimate the driver node. For medium to high
noise levels, the `sad' emerged as the driver node.

\begin{figure}[htbp]
\centering
\includegraphics[width=.9\linewidth]{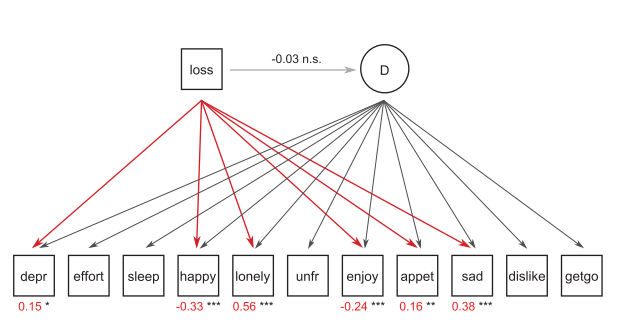}
\caption{\label{fig:fried_association}Main results from Friend and colleagues  \cite{Fried2015}. The network represents the output from a Multiple Indicators Multiple Causes (MIMIC) model. The red lines indicate significant direct effects of spousal loss on Center for Epidemological Studies Depression Scale(CES-D items); standardized estimates of these affects are represented in red below the symptoms. There was no significant loading of loss on the latent factor \(D\). For more info see  \cite{Fried2015}.}
\end{figure}

In the original study, the bereavement  score was most affected by `lonely', and
showed   weak   negative   associations   with  `happy'   and   `effort'   (fig.
\ref{fig:fried_association} adopted  from \cite{Fried2015}). Consequently,  it seems
that medium  to high thermal  noise is most  congruent with the  original study.
Fried  and  colleagues postulated  that  `lonely'  was  the gateway  from  which
information spreads through the network, i.e. bereavement was embodied mainly by
`loneliness' which then  percolated its effect to the other  symptoms. Since the
nature of  the data was  cross-sectional, the  comparison with the  results from
this study relies  on the assumption that binary dynamics  are representative of
the absence and presence of psychological symptoms. If correct, the results from
this  study  give   a  causal  perspective  on  the   associative  results  from
\cite{Fried2015}. The results from this  study postulate that `depr', `lonely' and
`sad' have similar causal effect for moderate to high thermal noise.

It is important to emphasize that a quantification is given in terms of absolute
effect  size and  not directed  effects. This  means that  nudging for  instance
`sleep' has  some effect \(X\)  on the  psycho-symptom network, in  what direction
that  effect  is, or  whether  it  has a  positive  or  negative effect  on  the
bereavement score /  cognitive load of the  patient is not clear,  and should be
the subject of future studies.

As  a final  note, the  field of  psychometrics is  concerned with  relating how
observables (e.g. behavior, responses on questionnaires, \textit{etc}) relate to
theoretical cognitive  constructs such  as intelligence  or mental  disorders. A
common approach in  understanding high level phenomena such as  depression is to
use a latent variable model, i.e. assuming  some high abstract feature to be the
cause  of the  observables  (or vice  versa). Only  recently  has this  paradigm
shifted   from  a   latent  variable   model   to  a   network  based   approach
\cite{Waldorp2011,Epskamp2018,Borsboom2011}.  Marsman   \textit{et  al.}  recently
reconciled these two  approached by showing statistical  equivalence between the
Ising  model  and  canonically  used latent  variable  models  in  psychometrics
\cite{Marsman2018}. The two approaches thus  highlight different aspects in theory
building; measurement  invariance and  correlation structure may  be interesting
from a common  cause approach but not  from a network perspective  which is more
interested in dynamical aspects of the  system. Both approaches, however, aid in
highlighting different aspects of the psychological constructs.

\newpage
\subsection{Code manual}
\label{sec:org6e4c208}
Accompanying this paper, I developed  a general framework for analyzing discrete
systems using IMI. The code is written  in python 3.7.2 and uses cython 0.28.2
for c/c++ level performance. The code  is freely available on \codebase includes
the  latest build  instructions.hat  follows here  is a  brief  overview of  the
framework.

\subsection{Structural methods}
\label{sec:org9e3ed34}
Network analysis  has traditionally resulted  in analyzing the structure  of the
network. A  fundamental concept within network  science is centrality, and  how to
measure the  centrality of nodes has  become an essential part  of understanding
networked systems  such as social  networks, the internet,  biological networks,
traffic and  ecological networks. At  its core, a centrality  measure quantifies
the `importance' of a node based  on some structural property. It allows ranking
nodes based on a real-valued function.

There is, however, a long-standing debate concerning what centrality metrics
actually measure for networked systems
\cite{Bringmann2018,Borgatti2005,Borgatti2006,Sikic2013} . From a network
theoretical perspective most centrality measures, e.g. betweenness, closeness,
eigenvector and degree centrality, essentially classify the `walk structure' of
a network \cite{Borgatti2005,Borgatti2006}. A walk from node \(i\) to node \(j\) is a
sequence of adjacent nodes that begins with \(i\) and ends with \(j\). The structure
of walks can be divided along different criteria. For example a trail is a walk
in which no edge (i.e. adjacent pair of nodes) is repeated. In contrast, a path
is a trail in which no node is visited more than once. Similarly, one could
define a walk structure by only using the shortest path from one node to
another, or by using random movements between nodes (random walks).

Alternatively,   from  a   complex  systems   perspective,  centrality   metrics
\textit{implicitly  assume}  dynamics  on  the  network  structure.  Betweenness
centrality for example, computes centrality based on  how often a node acts as a
bridge along the shortest path between two  other nodes. If one assumes that the
network has  dynamics \(D\) where  information between nodes follows  the shortest
path, this  metric may be  a valid description  to use and  identify dynamically
important nodes.

In  the best  case,  a centrality  metric is  fully  predictive for  identifying
important nodes  a complex  system. Consequently, the  centrality metric  can be
used to  understand the  system. However,  an issue with  the use  of centrality
metrics  is determining  which centrality  metric to  use. Consider  for example
figure \ref{fig:methods1} A; different centrality  metrics can identify different
nodes  as most  central.  This has  lead  to the  common  observation that  some
centrality measures  can `get  it wrong'  when the aim  is to  predict dynamical
important  structure in  networked systems.  Additionally, the  ranking produced
through some  centrality metric does  not quantify inter-rank  differences. This
potentially leads  to underestimation  of nodal influence  when used  in dynamic
context \cite{Sikic2013}.

We will show how centrality measures  have no meaningful prediction power of the
most causal  node in  nodes dictated  by the  Gibbs measure.  We are  aware that
centrality measures  do not embody  the full  extent of what  structural methods
embody,  or what  network  science in  particular has  to  offer. However,  many
structural  methods share  the common  characteristics listed  above, i.e.  they
quantify the walk structure  of a network. For our analysis,  we used the weighted
variants of  degree centrality, betweenness centrality,  information centrality,
and eigenvector centrality. What follows is a brief description of commonly used
centrality metrics.

\subsubsection{Degree centrality}
\label{sec:org541063a}
Degree centrality is the best-known  measure of all the centrality measures.
It is  often thought that  degree centrality  is indicative for  the dynamic
importance of a  node. This intuition is  based on the concept  of flow: the
more connection a node has, the more interaction potential that node has and
therefore  the more  important a  node must  be. Freeman  defined centrality
measure as  the count  of the  number of  edges incident  upon a  given node
\cite{Freeman1979}:

\begin{equation}
    \label{eq:degree}
    c_i^{\textnormal{deg}} = \sum_{j} a_{ij}
\end{equation}
where \(a_{ij}\) is the row/column of node \(i\) in the adjacency matrix \(A\) of the network. Please note that the entries \(a_{ij}\) are weighted and not binary.
\subsubsection{Betweenness centrality}
\label{sec:org23f41df}
Betweenness centrality  quantifies the number of  times a node acts  as a bridge
along the shortest path between two other  nodes. It was introduced as a measure
for quantifying the control of communication  among humans in social networks by
Freeman  \cite{Freeman1979}. Nodes  that have  a high  probability to  occur on  a
randomly chosen shortest  path between two randomly chosen vertices  have a high
betweenness. Formally, this can be written as:

\begin{equation}
\label{eq:betweeness}
\begin{split}
\bet_i = \sum_{j, k} \frac{\sigma(j, k \vert i)}{\sigma(j, k)}
\end{split}
\end{equation}

where \(\sigma(j, k)\) represents the number of shortest paths between node \(j\) and \(k\), and \(\sigma(j, k \vert i)\) is the subset that goes through node \(i\). We use the normalized version of betweenness that divides the  betweenness score by the number of pairs of vertices (not including node \(i\));

\begin{equation}
    \begin{split}
    \label{eq:betweenness}
    \bet_i &= \frac{1}{Z}\sum_{j, k} \frac{\sigma(j, k \vert i)}{\sigma(j, k)}\\\
    Z  &= \frac{((n - 1) (n - 2))}{2}
    \end{split}
    \end{equation}

\subsubsection{Closeness centrality}
\label{sec:orgdfbe03b}
Closeness  centrality is  defined as  the reciprocal  sum of  the length  of the
shortest paths between  the node \(i\) and  all other nodes in the  network \(j \in
N\):

\begin{equation}
    \close_i= \frac{1}{\sum_j^N d(i, j)}.
    \label{eq:closeness}
\end{equation}
From a complex system perspective it assumes that information is transferred
along its  shortest paths. A  node with short  distance to many  other nodes
will  be able  to quickly  transfer its  information to  other nodes  in the
network.

\subsubsection{Eigenvector centrality}
\label{sec:org1c40770}
Eigenvector  centrality is  the most  difficult  centrality measure  to give  an
intuitive  feeling  for. Where  \(A\)  is  the  adjacency  matrix of  the  system,
eigenvector centrality of node \(i\) is defined as:

\begin{equation}
\label{eq:eigenvector}
\begin{split}
    \ev_{i} = \frac{1}{\lambda} \sum_{j} a_{ij} x_{j} \leftrightarrow Ax = \lambda x
\end{split}
\end{equation}

For  any  square  matrix  of  rank  \(n\),  the  matrix  will  have  at  most  \(n\)
eigenvector-eigenvalues  pairs. A  common choice  for eigenvector  centrality is
motivated by The Perron-Frobenius theorem, and involves choosing the eigenvector
\(x\) with the largest  eigenvalue \(\lambda\) \cite{Debye1918,Frobenius1912}. This
has the  desired property  that if  \(A\) is irreducible,  or equivalently  if the
network  is strongly  connected,  that  the eigenvector  \(x\)  is  both unique  and
positive.

The sign and size  of the eigenvalue are important for  the relation between the
value  and importance  of  a  node. In  linear  differential equations  negative
eigenvalues  correspond to  non-oscillatory exponentially  stable solutions.  In
contrast,  in   difference  equations  it  indicates   an  oscillatory  behavior.
Geometrically speaking,  negative eigenvector  embodies a  linear transformation
across some axis.

Intuitively speaking, eigenvector  centrality quantifies the
influence of  a node  in the  network. It  assigns relatives
scores to all nodes in the network based on the concept that
connections  to high-scoring  nodes contribute  more to  the
score  of the  node in  question than  equal connections  to
low-scoring nodes. A high eigenvector score implies that the
node is connected  to many other nodes  that themselves have
high  scores.  Google  PageRank   and  Katz  centrality  are
variants  of  eigenvector centrality  \cite{Langville2005}.  A
node with  high eigenvector centrality is  not necessarily a
node that  has many  connection (incoming or  outgoing). For
example a node may have  a high eigenvector centrality if it
has few connections, but  those connections are connected to
nodes that are of high importance.

\subsection{Bootstrap distributions}
\label{sec:org603bd1b}
A total of \(\num{10000}\)  bootstrap samples were
conducted with  replacement. For each  nodal bootstrap distribution  a gaussian
kernel  density was  estimated. The  node with  the highest  causal impact  was
chosen as  the initial driver  set \(\Lambda\).  Then the overlap  \(\phi\) between
this  driver  node  and  the  remaining   nodes  in  the  system  was  computed
iteratively. We  considered the overlap \(\phi  = 0.5\) or higher  was sufficient
for  the  node was  to  be  considered causally  similar  to  the driver  node.
Therefore, the proposed node  will be included in the driver  node set if \(\phi
 \geq 0.5\).

\subsubsection{Numerical procedure}
\label{sec:orgdd91c2c}
\begin{listing}[H]
\begin{minted}[frame=lines,linenos=true]{python}
def bootstrapDrivers(D, N, M, phi=0.5) -> dict:
   """
   Determines driver-nodes through the  boostrap distribution It consists as an
   iterative procedure  that takes  the max  value in the  data, then  a normal
   distribution is  fitted an tested  with overlap for  the other nodes  in the
   data at `phi`

   Parameters
   ----------
   D: np.array
       2d matrix consisting of size n_features x n_observations
   N: int
       Total number of samples used for bootstrap
   M: int,
   number of draws for each of each bootstrap sample
   phi : float
       Overlap between distribution.

   Returns
   ----------
       dict of driver-nodes under overlap `phi`
   """
   # part of standard library python
   from statistics import NormalDist
   import numpy as np

   # generate subsamples
   n_features, n_obs = D.shape
   D_bar = np.zeros((n_features, N))
   # sample from observations
   for var in range(n_features):
       D_bar[var] = np.random.choice(D[var], size=(N, M)).mean(1)

   # create bootstrap distribution
   driver = D_bar.mean(1).argmax()
   driverDist = NormalDist().from_samples(D_bar[driver])

   # other nodes to consider
   drivers = {} # Lambda
   options = np.arange(n_features)
   for var in options:
       # fit distribution
       otherDist = NormalDist().from_samples(D_bar[var])
       # compute overlap
       overlap = driverDist.overlap(otherDist)
       if overlap > phi:
           drivers[var] = (otherDist.mean, otherDist.variance)
   return drivers
\end{minted}
\caption{\label{alg:driver_set}Driver node detection algorithm in python.}
\end{listing}

\newpage

\subsubsection{Erdös-Rényi networks}
\label{sec:org967d812}
Bootstrap  distribution kernel  density  estimates for  Erdös-Rényi networks.  X
indicates  area  under  the  curve  for  mutual  information  or  causal  impact
respectively. Note  \(x\) here  refers to  the input variable  which are  the area
under curves either for integrated mutual information (\(\eta = 0\)) or for causal
impact \(\eta > 0\).

\begin{figure}[H]
\centering
\includegraphics[width=.9\linewidth]{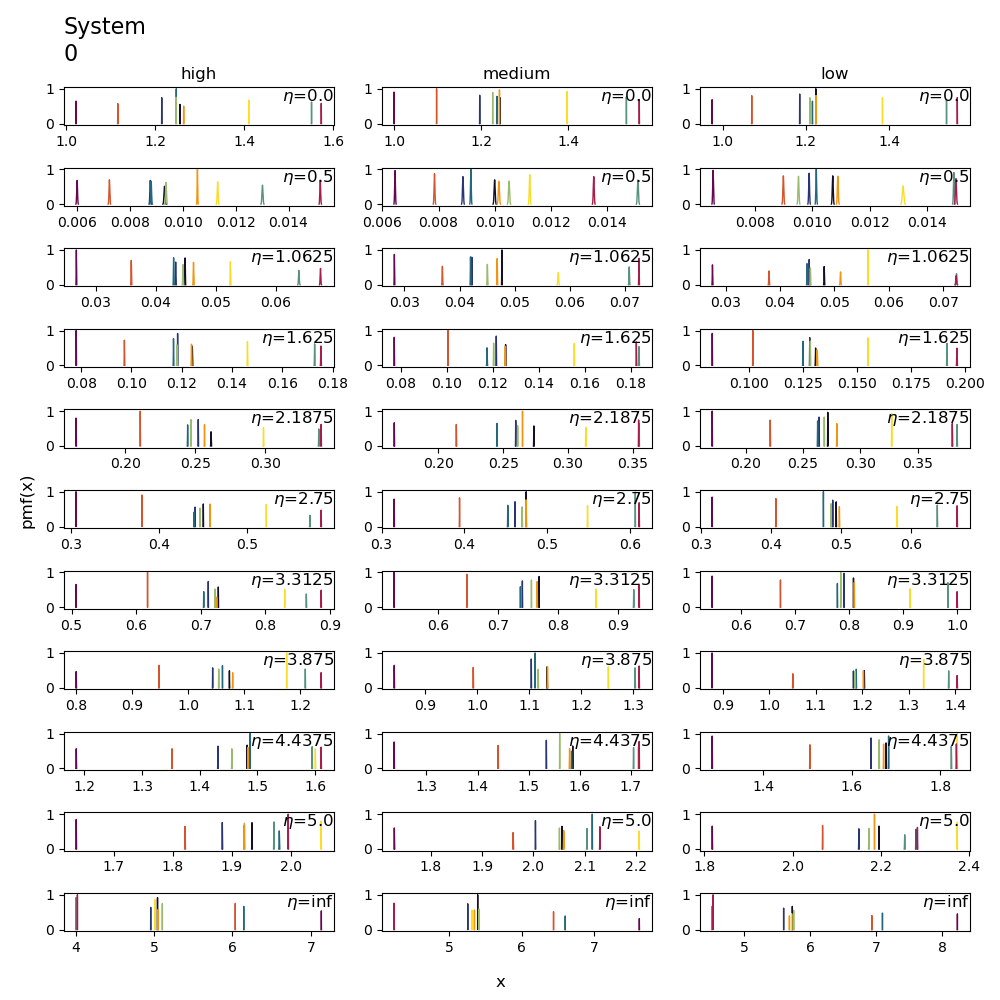}
\label{fig:bootstrap_distribution}
\end{figure}

\begin{figure}[H]
\centering
\includegraphics[width=.9\linewidth]{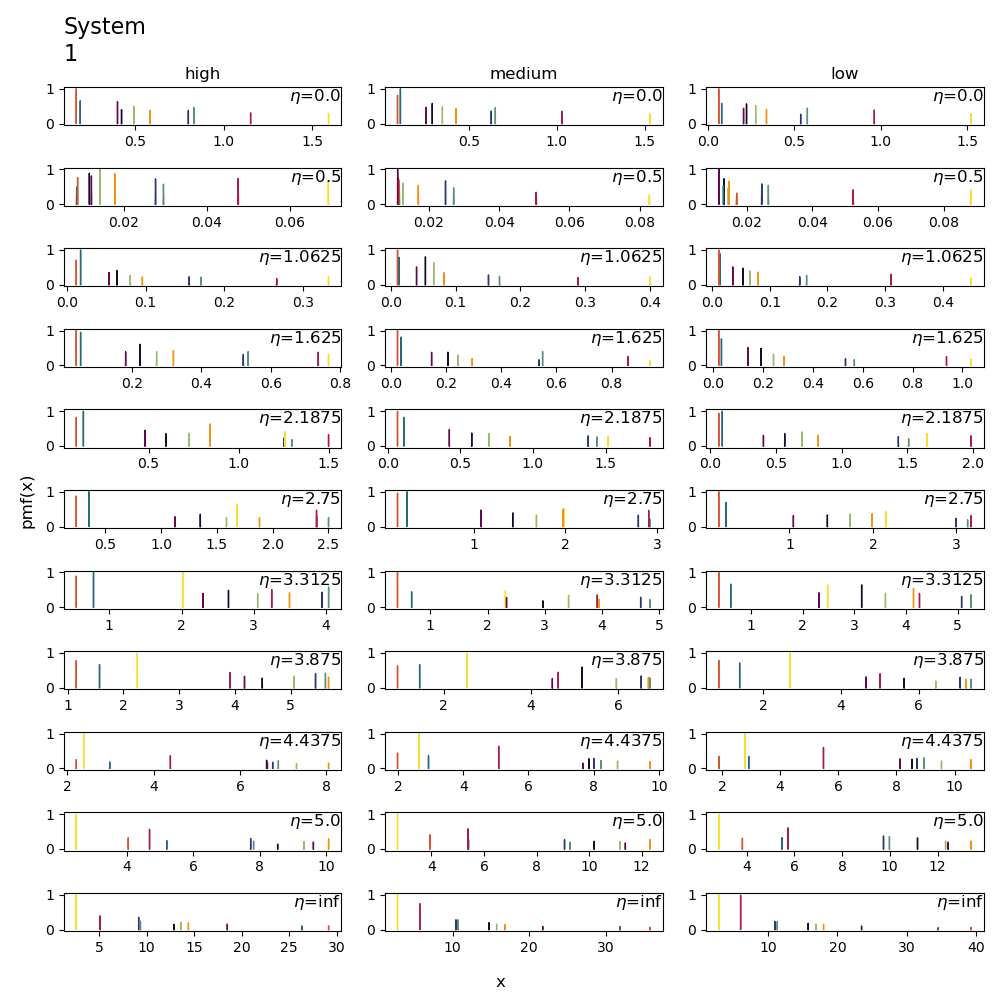}
\label{fig:bootstrap_distribution}
\end{figure}

\begin{figure}[H]
\centering
\includegraphics[width=.9\linewidth]{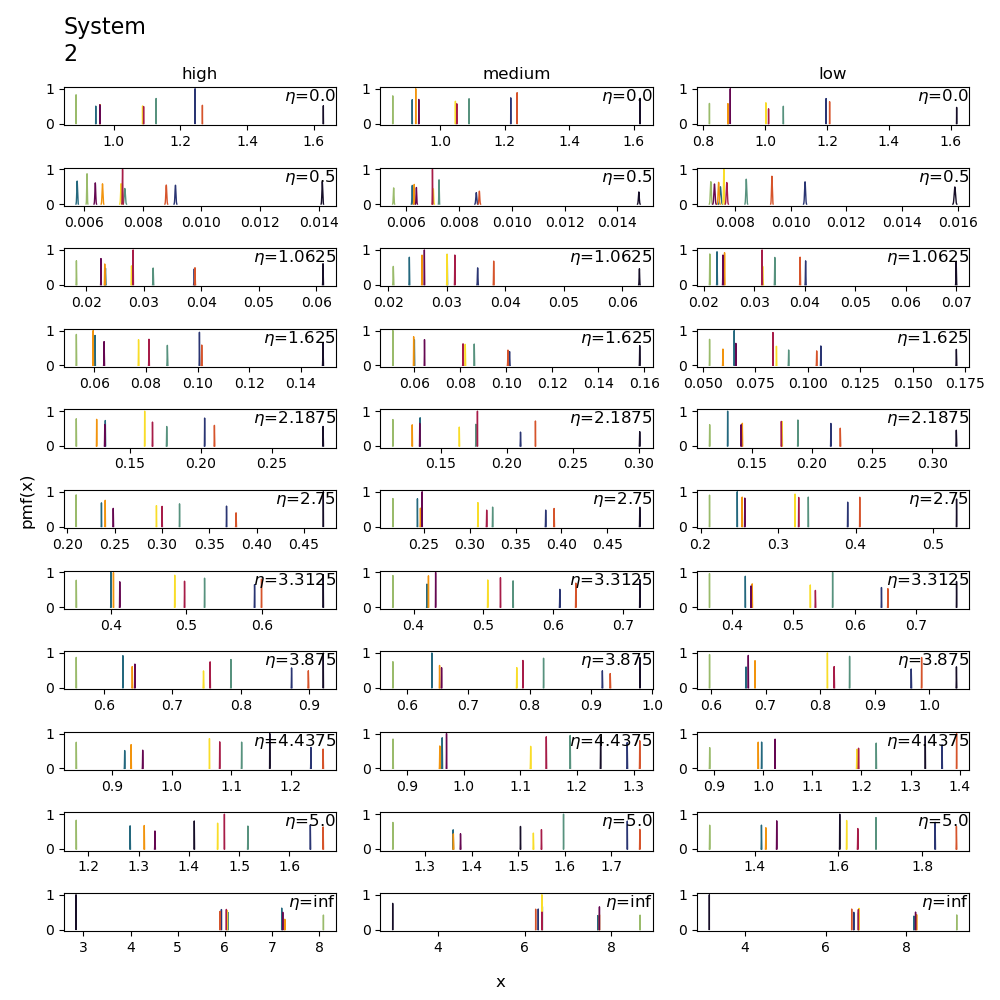}
\label{fig:bootstrap_distribution}
\end{figure}

\begin{figure}[H]
\centering
\includegraphics[width=.9\linewidth]{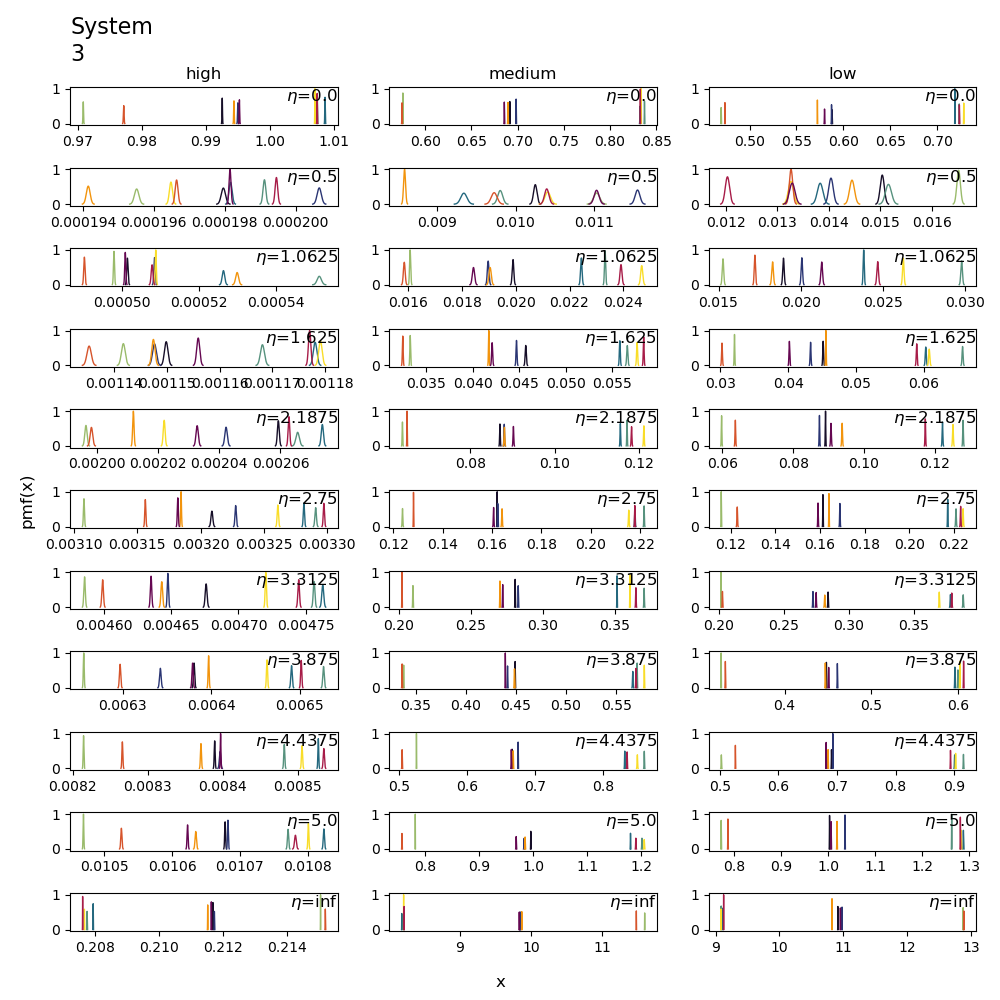}
\label{fig:bootstrap_distribution}
\end{figure}

\begin{figure}[H]
\centering
\includegraphics[width=.9\linewidth]{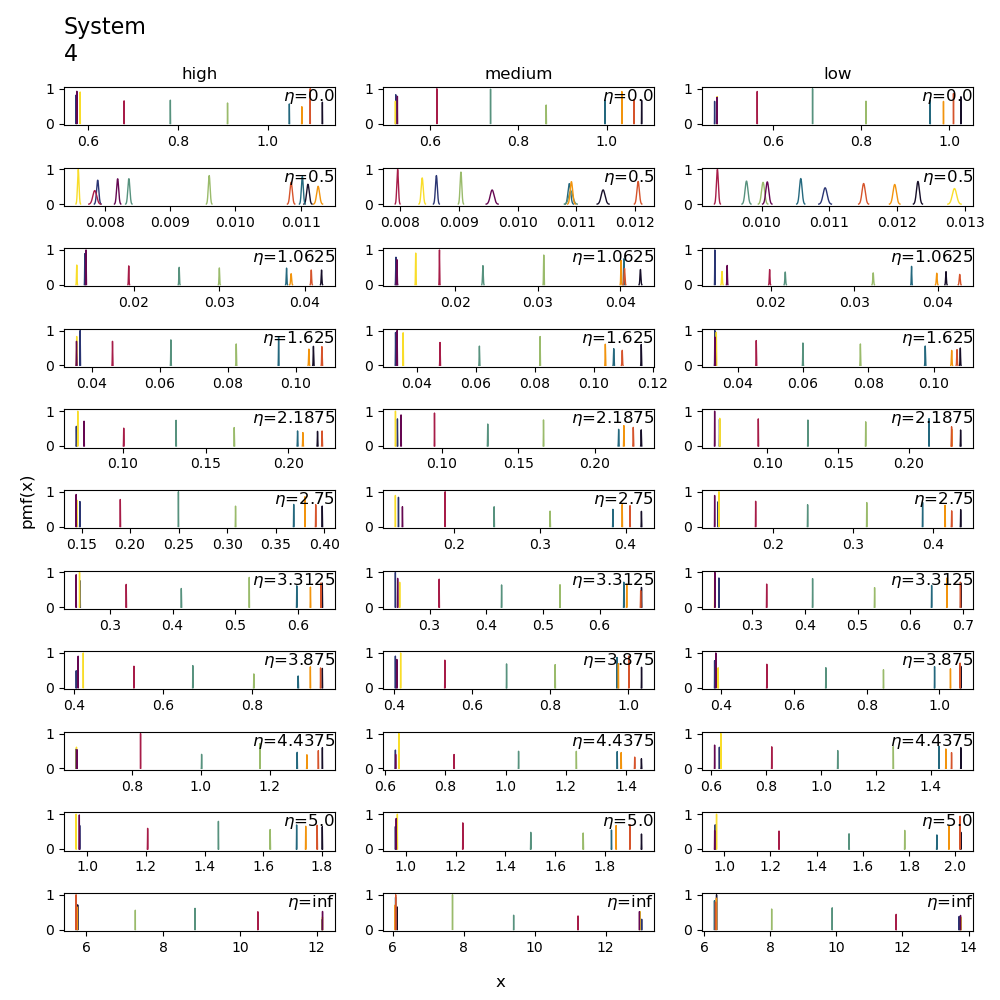}
\label{fig:bootstrap_distribution}
\end{figure}

\begin{figure}[H]
\centering
\includegraphics[width=.9\linewidth]{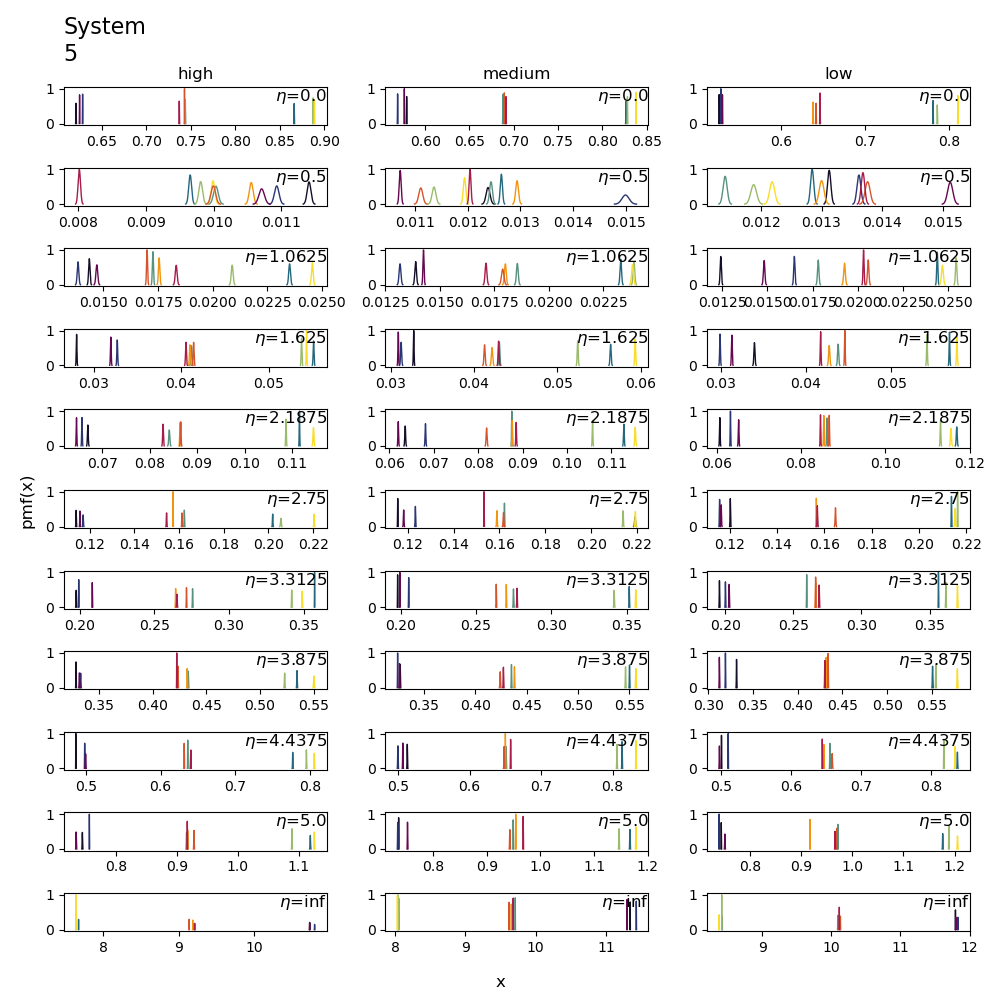}
\label{fig:bootstrap_distribution}
\end{figure}

\begin{figure}[H]
\centering
\includegraphics[width=.9\linewidth]{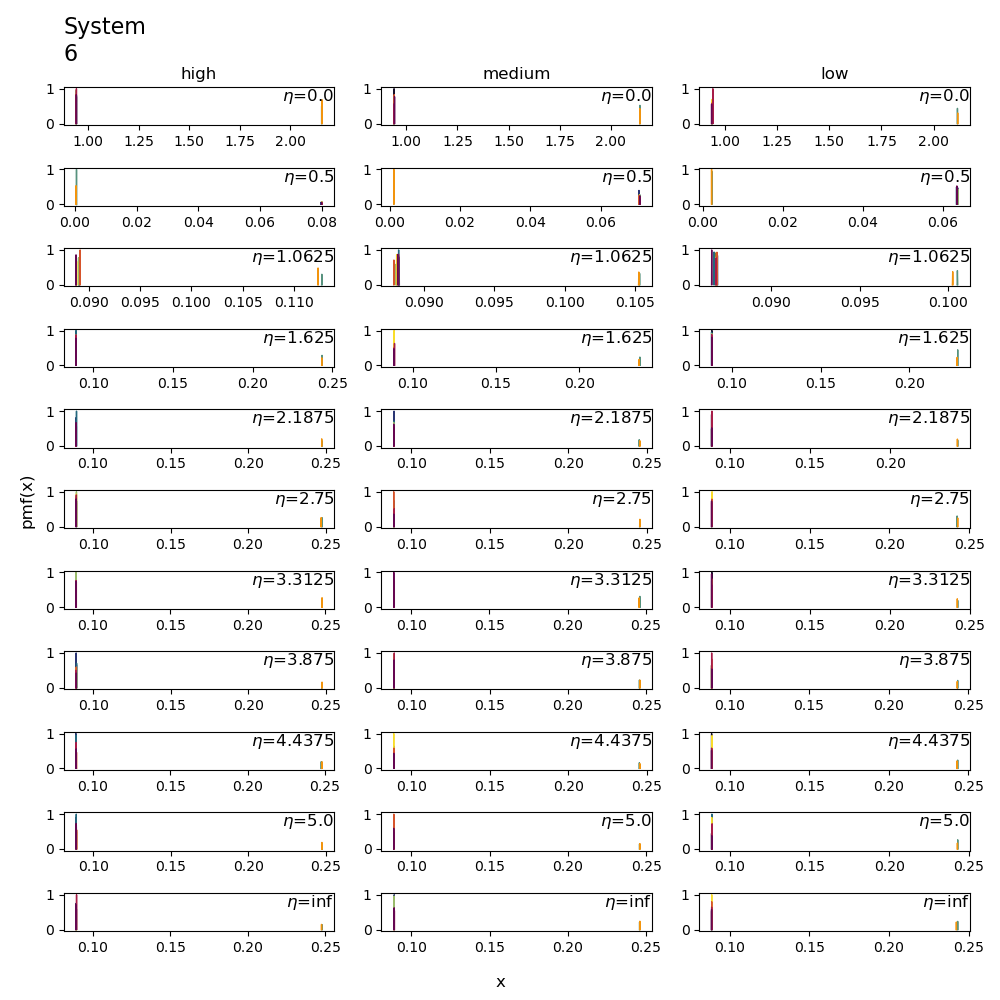}
\label{fig:bootstrap_distribution}
\end{figure}

\begin{figure}[H]
\centering
\includegraphics[width=.9\linewidth]{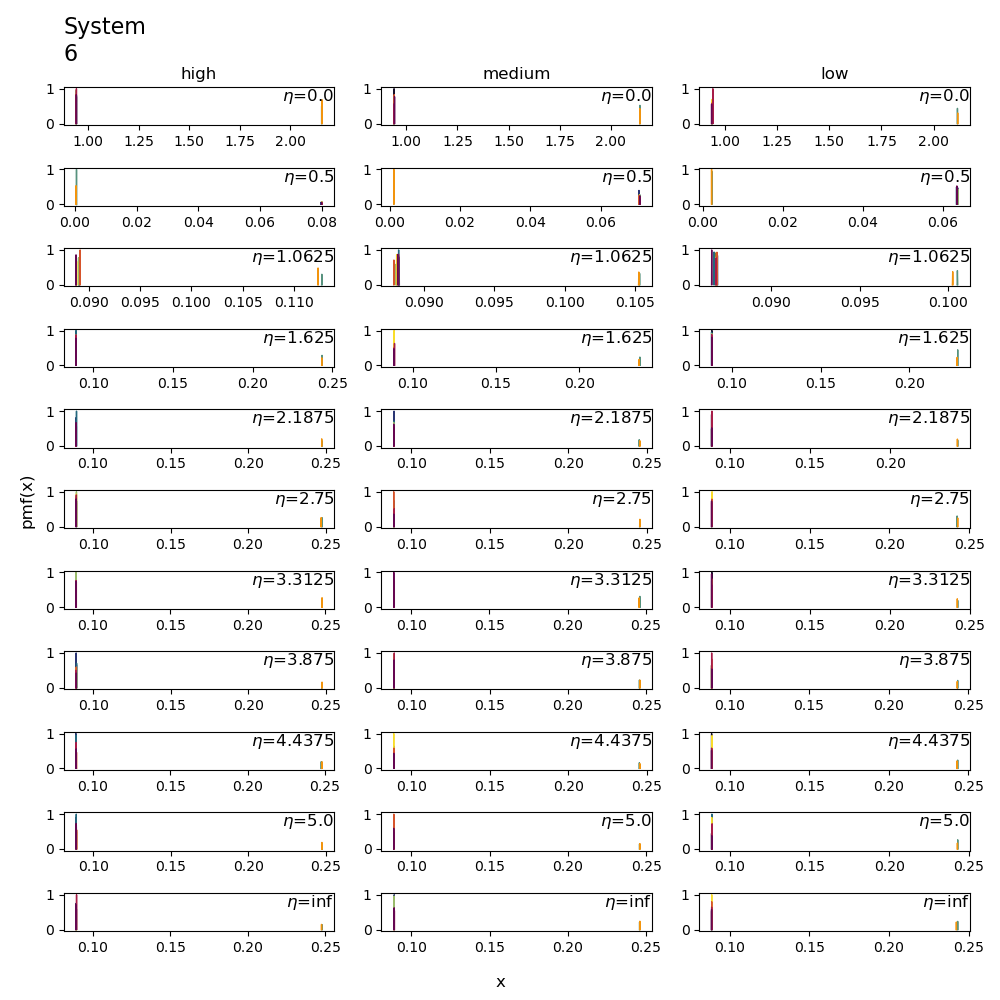}
\label{fig:bootstrap_distribution}
\end{figure}

\begin{figure}[H]
\centering
\includegraphics[width=.9\linewidth]{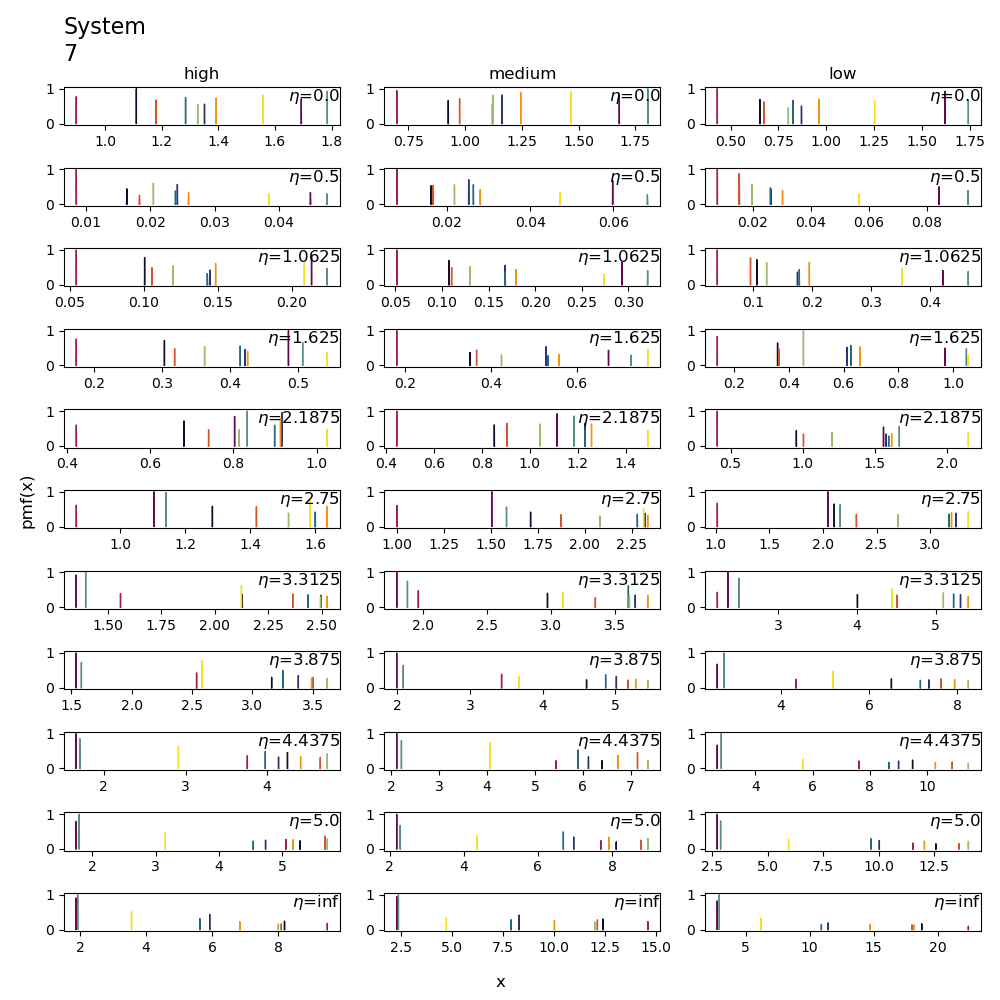}
\label{fig:bootstrap_distribution}
\end{figure}

\begin{figure}[H]
\centering
\includegraphics[width=.9\linewidth]{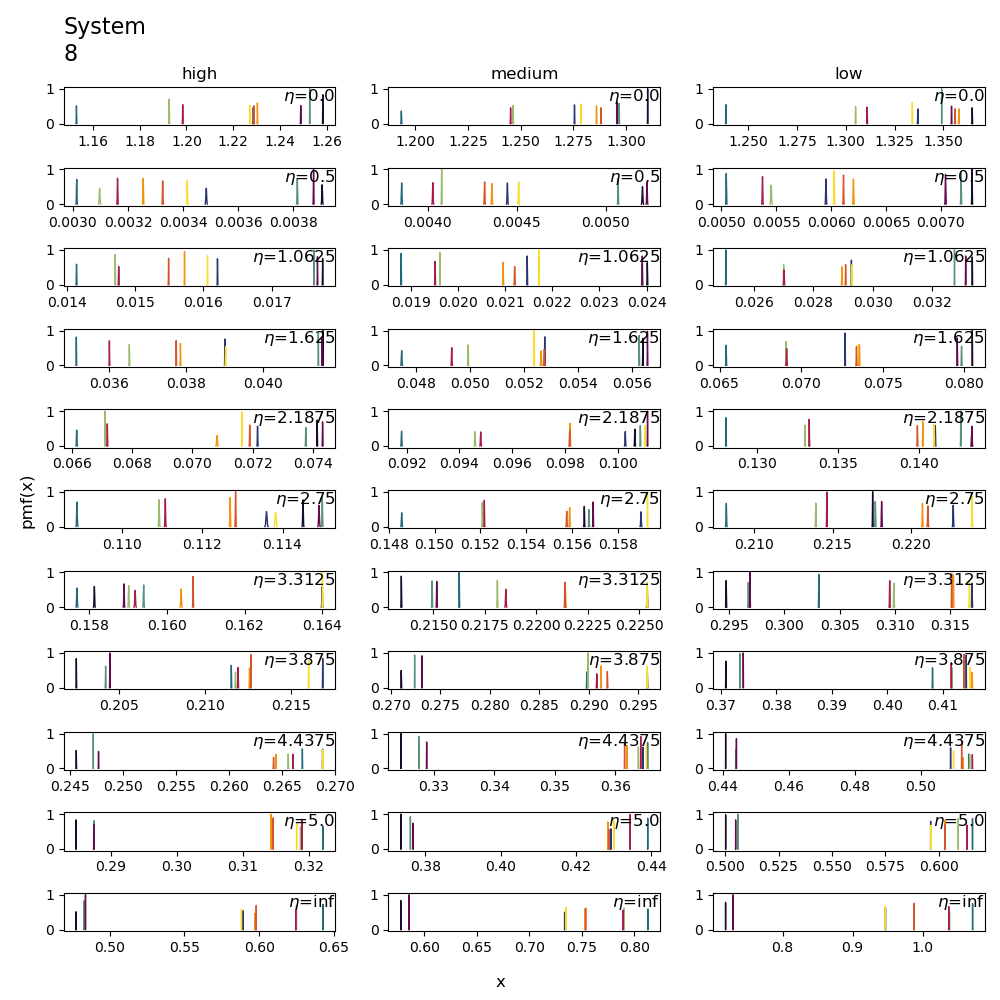}
\label{fig:bootstrap_distribution}
\end{figure}

\begin{figure}[H]
\centering
\includegraphics[width=.9\linewidth]{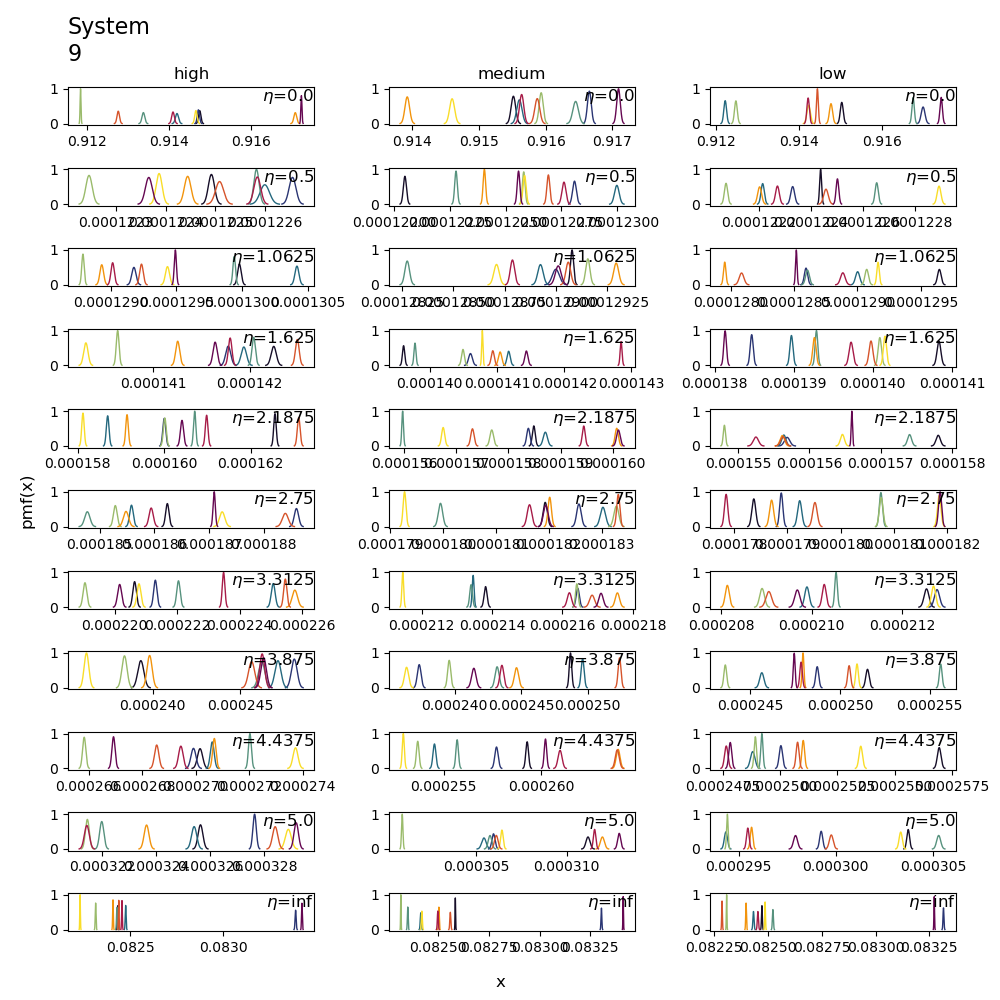}
\label{fig:bootstrap_distribution}
\end{figure}

\begin{figure}[H]
\centering
\includegraphics[width=.9\linewidth]{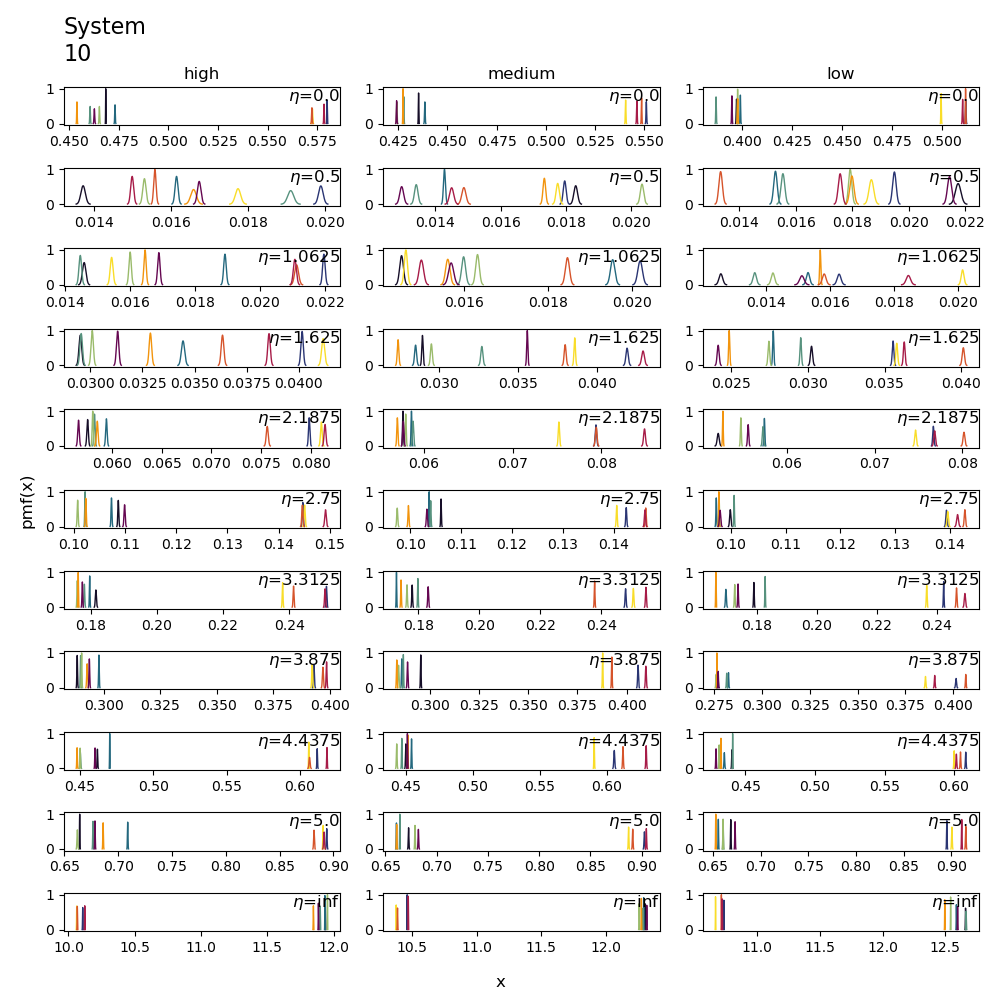}
\label{fig:bootstrap_distribution}
\end{figure}

\begin{figure}[H]
\centering
\includegraphics[width=.9\linewidth]{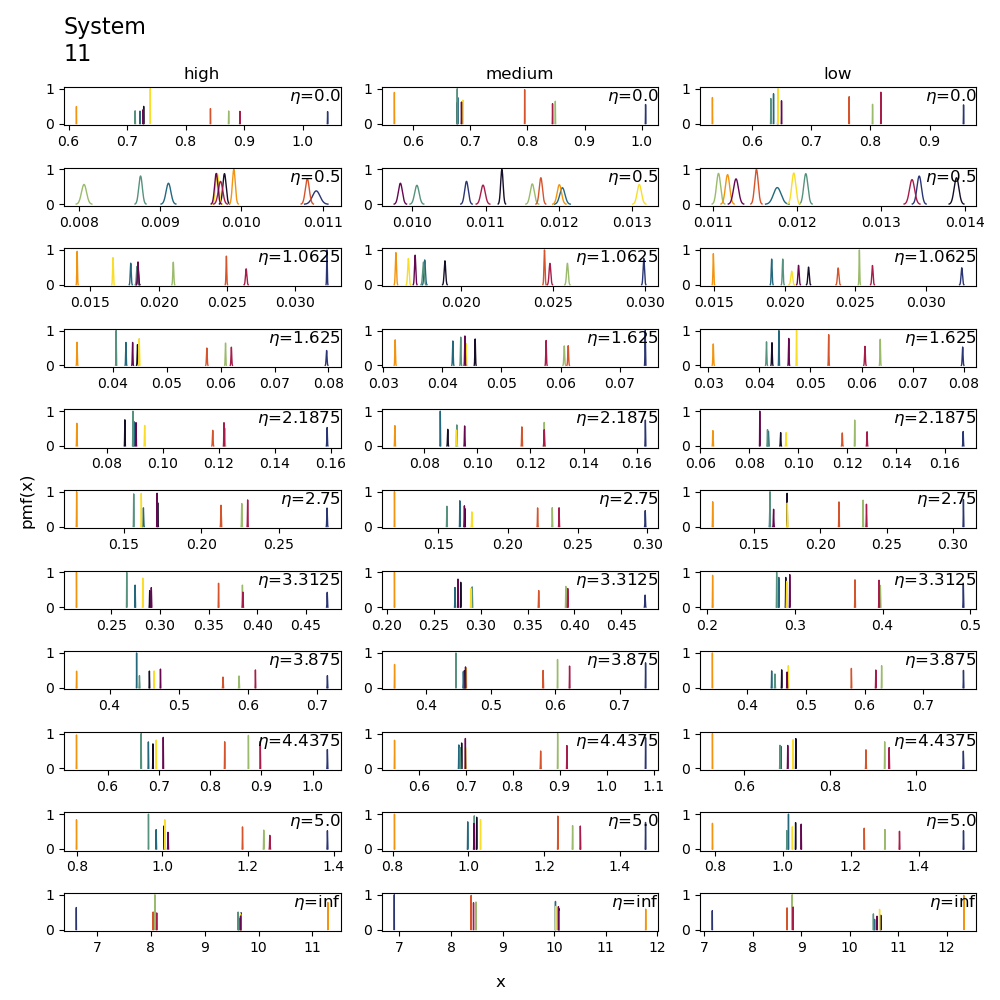}
\label{fig:bootstrap_distribution}
\end{figure}

\begin{figure}[H]
\centering
\includegraphics[width=.9\linewidth]{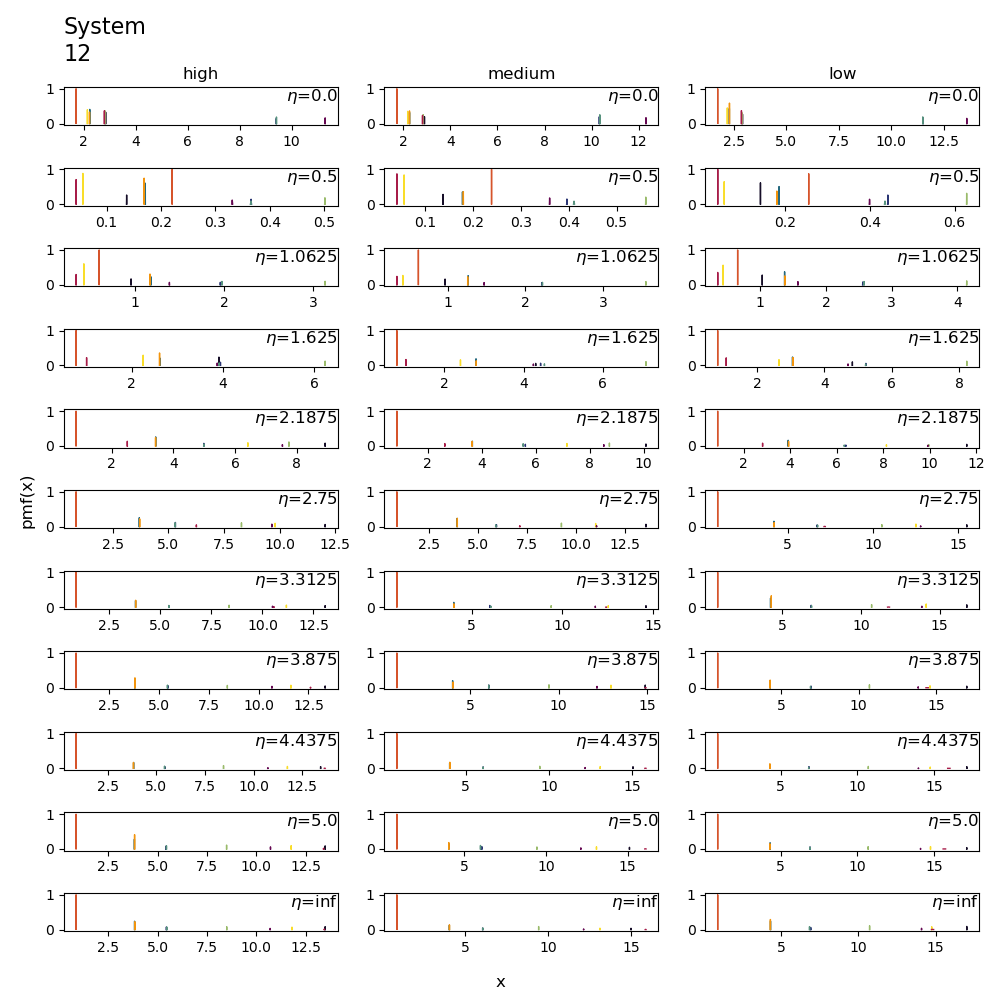}
\label{fig:bootstrap_distribution}
\end{figure}

\begin{figure}[H]
\centering
\includegraphics[width=.9\linewidth]{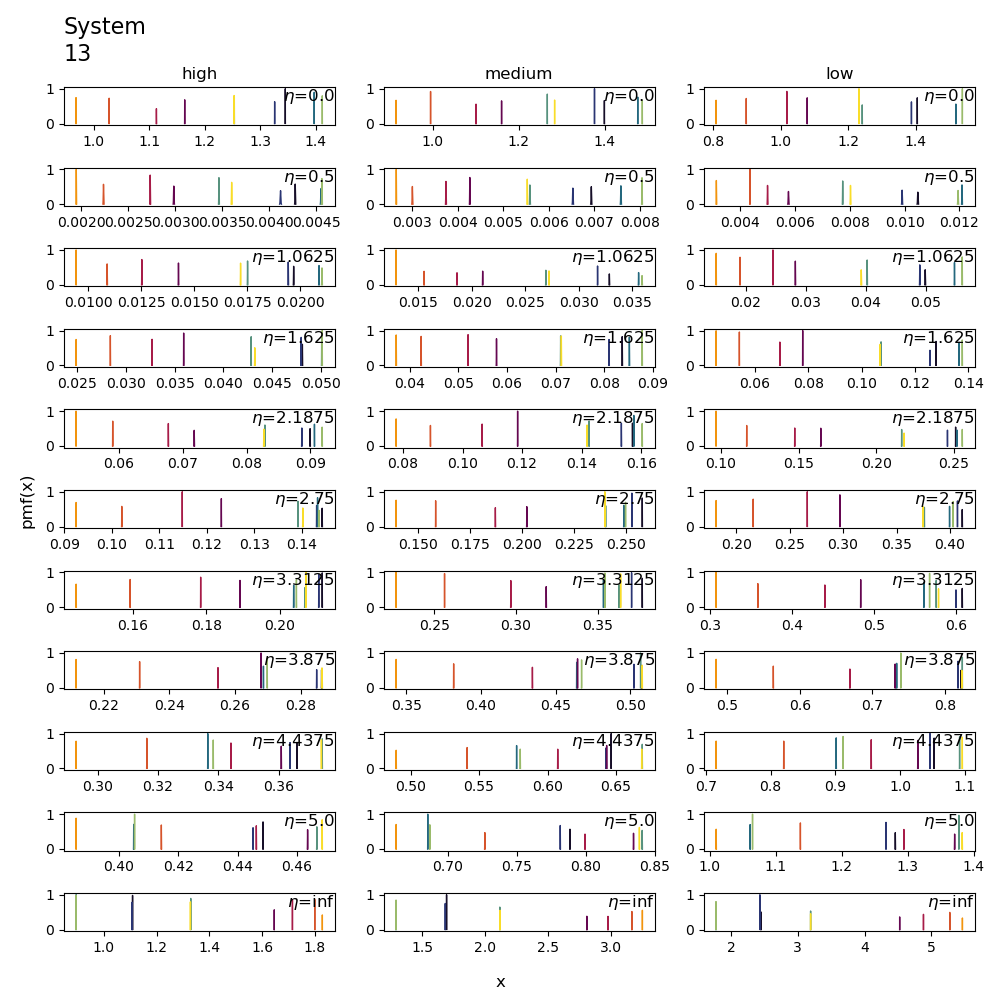}
\label{fig:bootstrap_distribution}
\end{figure}

\begin{figure}[H]
\centering
\includegraphics[width=.9\linewidth]{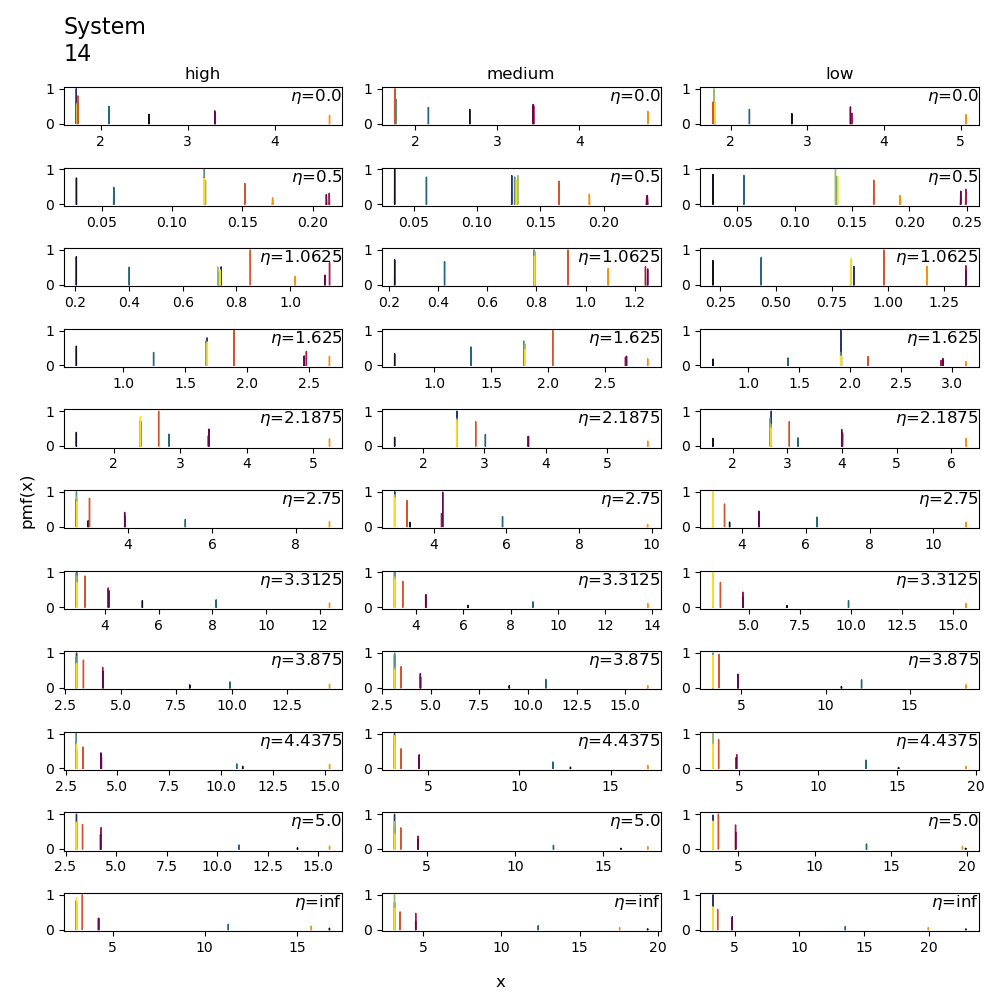}
\label{fig:bootstrap_distribution}
\end{figure}

\begin{figure}[H]
\centering
\includegraphics[width=.9\linewidth]{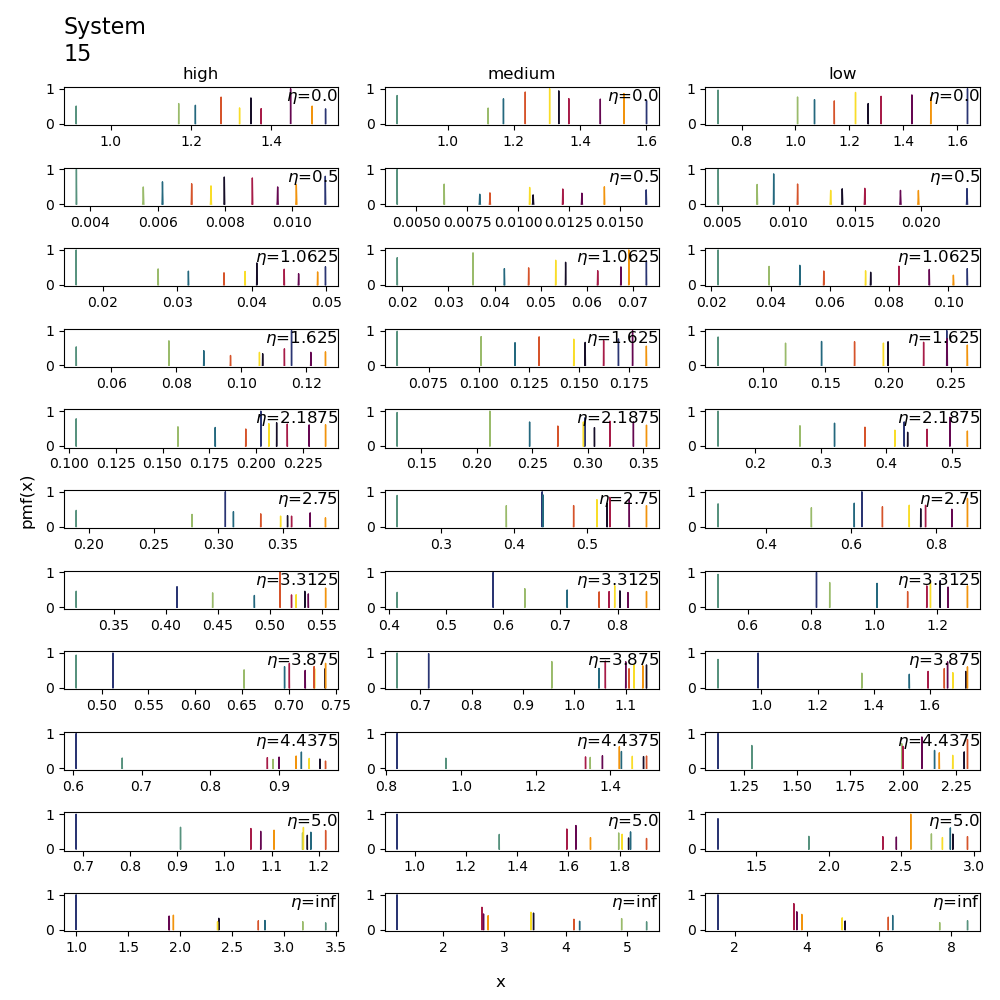}
\label{fig:bootstrap_distribution}
\end{figure}

\subsubsection{Psycho networks}
\label{sec:org65e0376}
Note \(x\)  here refers  to the  input variable  which are  the area  under curves
either for integrated mutual information (\(\eta = 0\)) or for causal impact \(\eta
> 0\).

\begin{figure}[htbp]
\centering
\includegraphics[width=.9\linewidth]{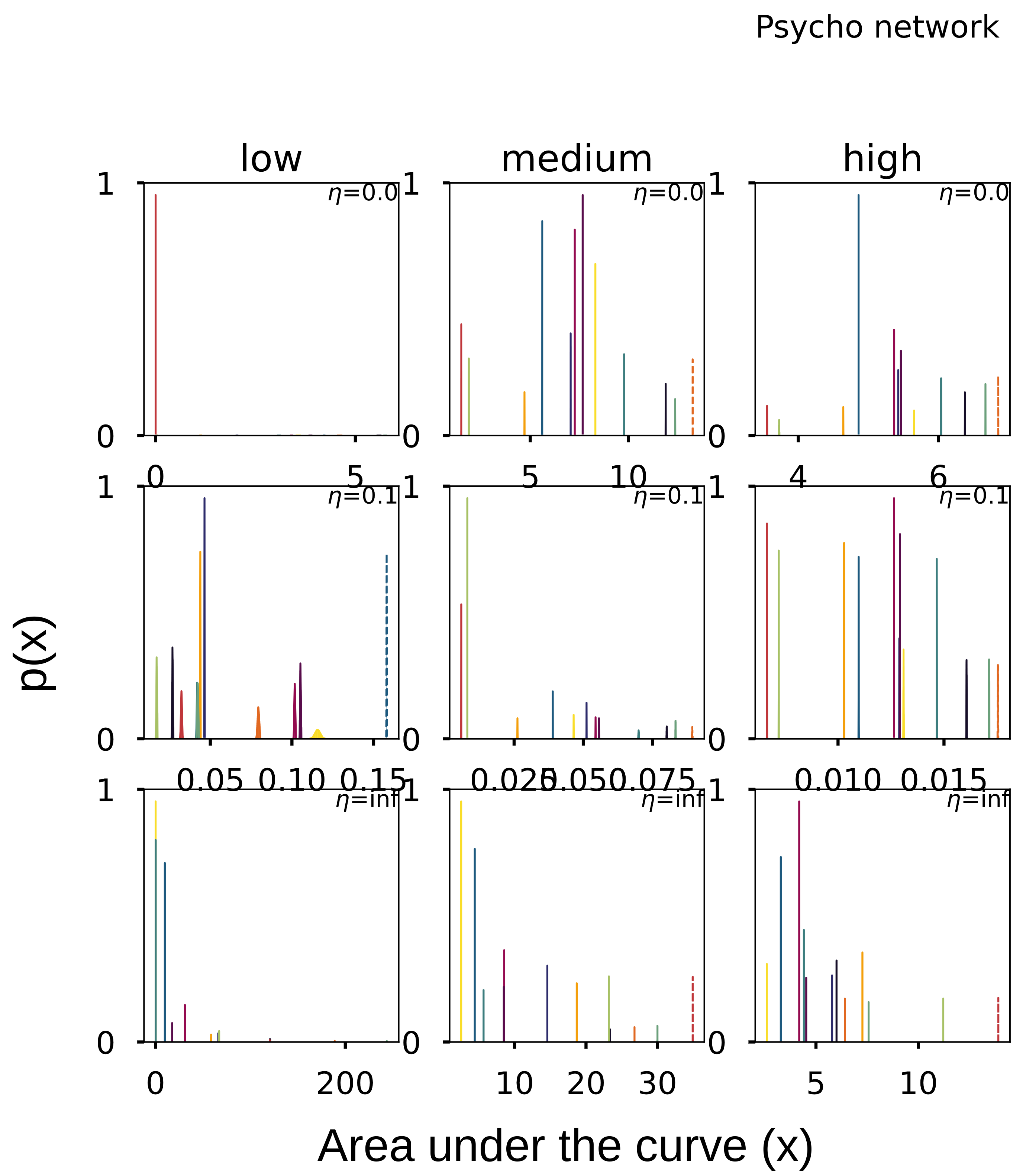}
\caption{\label{fig:label}Bootstrap distribution for psychosymptom network. Driver node is indicated by a dotted line.}
\end{figure}

\newpage
\end{document}